\documentclass[twocolumn,showpacs,preprintnumbers,
               superscriptaddress]{revtex4}
\usepackage{graphicx}

\def\lsim{\raise0.3ex\hbox{$<$\kern-0.75em\raise-1.1ex\hbox{$\sim$}}}
\def\gsim{\raise0.3ex\hbox{$>$\kern-0.75em\raise-1.1ex\hbox{$\sim$}}}

\begin{document}

\preprint{UTCCP-P-135}


\title{Two flavors of dynamical quarks on anisotropic lattices}


\newcommand{\Tsukuba}%
{Institute of Physics, University of Tsukuba, 
 Tsukuba, Ibaraki 305-8571, Japan}

\newcommand{\CCP}%
{Center for Computational Physics, University of Tsukuba, 
 Tsukuba, Ibaraki 305-8577, Japan}

\newcommand{\ICRR}%
{Institute for Cosmic Ray Research, University of Tokyo, 
 Kashiwa, Chiba 277-8582, Japan}

\newcommand{\Hiroshima}%
{Department of Physics, Hiroshima University, 
 Higashi-Hiroshima, Hiroshima 739-8526, Japan}

\newcommand{\KEK}%
{High Energy Accelerator Research Organization (KEK), 
 Tsukuba, Ibaraki 305-0801, Japan}

\author{T.~Umeda\footnote{
Address after Apr.\ 2003: YITP, Kyoto University, Kyoto 606-8502, Japan.}}
\affiliation{\CCP}

\author{S.~Aoki}
\affiliation{\Tsukuba}

\author{M.~Fukugita}
\affiliation{\ICRR}

\author{K.-I.~Ishikawa}
\affiliation{\CCP}

\author{N.~Ishizuka}
\affiliation{\CCP}
\affiliation{\Tsukuba}

\author{Y.~Iwasaki}
\affiliation{\CCP}
\affiliation{\Tsukuba}

\author{K.~Kanaya}
\affiliation{\Tsukuba}

\author{Y.~Kuramashi}
\affiliation{\KEK}

\author{V.I.~Lesk}
\affiliation{\CCP}

\author{Y.~Namekawa}
\affiliation{\Tsukuba}

\author{M.~Okawa}
\affiliation{\Hiroshima}
	
\author{Y.~Taniguchi}
\affiliation{\Tsukuba}

\author{A.~Ukawa}
\affiliation{\CCP}
\affiliation{\Tsukuba}

\author{T.~Yoshi\'{e}}
\affiliation{\CCP}
\affiliation{\Tsukuba}

\collaboration{CP-PACS Collaboration}
\noaffiliation

\date{\today}

\pacs{12.38.Gc}

\begin{abstract}
We report on our study of two-flavor full QCD on anisotropic
lattices using $O(a)$-improved Wilson quarks coupled with an
RG-improved glue.  
The bare gauge and quark anisotropies corresponding to the renormalized
anisotropy $\xi=a_s/a_t = 2$ are determined as functions of 
$\beta$ and $\kappa$, which covers the region of spatial lattice 
spacings $a_s\approx 0.28$--0.16 fm and 
$m_{PS}/m_V\approx 0.6$--0.9.
The calibrations of the bare anisotropies are performed with 
the Wilson loop and the meson dispersion relation at 4 lattice cutoffs
and 5--6 quark masses.
Using the calibration results we calculate the meson mass spectrum 
and the Sommer scale $r_0$.  We confirm that 
the values of $r_0$ calculated for 
the calibration using pseudo scalar and vector meson energy momentum 
dispersion relation coincide in the continuum limit within errors.
This work serves to lay ground toward studies of heavy quark systems and 
thermodynamics of QCD 
including the extraction of the equation of state in the continuum 
limit using Wilson-type quark actions. 

\end{abstract}

\maketitle

\section{Introduction}
 \label{sec:Introduction}

In spite of recent progress in computer technology and numerical algorithms, 
extraction of continuum properties from lattice QCD
remains to be challenging when dynamical quarks are included 
due to large computational demands. 
One of methods for alleviating the difficulty is to improve 
the lattice action for a faster approach to the continuum limit.
This has enabled us to carry out 
the first systematic extrapolation to the chiral and 
continuum limits for the light hadron spectrum 
\cite{cppacs00,cppacs02}. 

Another method which is effective for several quantities is to 
introduce a space-time anisotropy. 
In Ref.~\cite{namekawa01}, we have shown that using anisotropic lattices 
with a larger temporal cutoff is efficient for reducing lattice 
artifacts in thermal QCD, and have carried out the first well-controlled 
continuum extrapolation of the equation of state in quenched QCD.
In finite temperature QCD, anisotropic lattices have been employed 
in the quenched approximation also 
to study transport coefficients \cite{sakai00}, 
pole masses \cite{taro00,umeda01}, 
glueballs \cite{ishii02}, 
and spectral functions \cite{asakawa02,umeda02},
where anisotropy was introduced to obtain more data points 
for temporal correlation functions.
At zero temperature, anisotropic lattices have been employed to study 
charmonium states \cite{klassen98,chen01,okamoto02}, 
glueballs \cite{morningstar99}, 
heavy hybrids \cite{manke99,harada02}, 
and also the pion scattering length \cite{liu02}. 

In this paper, we calculate the anisotropy parameters for an improved full 
QCD action to contribute toward a systematic study of QCD 
with heavy quarks and at finite temperatures. 
The calculation of anisotropy parameters is not a simple task in full QCD 
because a couple of bare parameters have to be simultaneously adjusted to 
achieve a consistent renormalized anisotropy in physical observables for 
both quarks and gluons. 
This tuning of bare anisotropy parameters is called ``calibration'' 
\cite{hashimoto93}.
We study two-flavor full QCD with an RG-improved gauge
action and a clover-improved Wilson quark action, extending the combination
of improved actions adopted by the CP-PACS Collaboration to anisotropic 
lattices.
Carrying out simulations at several values of bare parameters, 
we perform the calibration to determine the bare anisotropy parameters 
for a given value of the renormalized anisotropy $\xi=a_s/a_t$ 
as functions of the gauge coupling and bare quark mass. 
We study the range of parameters corresponding to 
$a_s\approx 0.28$--0.16 fm for the spatial lattice spacing 
and $m_{PS}/m_V\approx 0.6$--0.9 
for the ratio of pseudoscalar and vector meson masses.
Based on our previous study of finite temperature QCD \cite{namekawa01}, 
we concentrate on the case $\xi=2$ in this paper.

Different choices of observables for the calibration will lead to 
$O(a)$ differences in the calibration results. 
We study this issue by comparing the results from two different 
observables 
--- pseudoscalar and vector meson dispersion relations.
We anticipate that the results of different calibrations will be useful 
for checking the stability of continuum extrapolations.
As a test of the idea, we also perform a continuum extrapolation of 
the Sommer scale at $\xi=2$, 
by interpolating our measurement results to the calibrated points. 

This paper is organized as follows:
We define our anisotropic lattice action in Sec.~\ref{sec:action}, and 
discuss our choice of $\xi=2$ and simulation parameters in 
Sec.~\ref{sec:parameter}.
The calibration procedure is described in Sec.~\ref{sec:calib_proc}.
Results of two calibrations for $\xi=2$, using pseudoscalar and vector 
meson dispersion relations, are summarized in Sec.~\ref{sec:calib_result}. 
Finally, in Sect.~\ref{sec:physics}, we interpolate the measurement 
results to $\xi=2$ to study basic properties of our lattices. 
We also test how the difference in the calibration affects 
physical observables.
Sec.~\ref{sec:conclusion} is devoted to conclusions and discussions.
An appendix is added to compare our calibration procedure 
with another method based on the ratio of screening and temporal masses.

\section{Anisotropic lattice action}
\label{sec:action}

We study full QCD with two flavors of degenerate light quarks.
On isotropic lattices, 
we have made a series of systematic studies adopting a clover-improved quark 
action coupled with an RG-improved glue 
\cite{cppacs00,cppacs02,cppacs00PT,cppacs01B,cppacs01EOS,cppacs01Topo,cppacs03Flv}.
In these studies, the clover coefficient $c_{SW}$ was set to the tadpole-
improved value using the plaquette in one-loop perturbation theory 
for the mean-field.  
This choice was based on the observations that the one-loop plaquette 
reproduces the actual plaquette expectation values within 8\% for the 
range of parameters studied, 
and that the resulting value of the clover coefficient agrees well 
with actual one-loop value of it. 
This action was shown to give both a good rotational symmetry of the heavy 
quark potential and a small scale violations in the light hadron spectra 
at moderate lattice spacings \cite{comparative}.
At zero temperature, these good properties enabled us to carry out first 
systematic chiral and continuum extrapolations of light hadron spectra and 
light quark masses \cite{cppacs00,cppacs02}.
At finite temperatures, this combination of actions was shown to reproduce 
the expected O(4) scaling around the two-flavor chiral transition point 
\cite{cppacs00PT,tsukuba97}, 
and was adopted in the first systematic calculation of the equation of 
state in lattice QCD with Wilson-type quarks \cite{cppacs01EOS}. 
Here, we extend the study to anisotropic lattices.

\subsection{RG-improved gauge action on anisotropic lattice}
\label{sec:RG}

On isotropic lattices, the RG-improved gauge action by Iwasaki \cite{iwasaki85}
consists of plaquettes and $1\times2$ rectangular loops. 
Extending it to anisotropic lattices, the general form of the action 
is given by
\begin{eqnarray}
&& S_G = \beta\left\{\frac{1}{\gamma_G}\sum_{x,i>j}
             \{c_0^s P_{ij}(x)+c_1^s (R_{ij}(x)+R_{ji}(x))\} \right.
\nonumber\\
&& \left. + \gamma_G\sum_{x,k}
             (c_0^t P_{k4}(x)
             +c_1^t R_{k4}(x)
             +c_2^t R_{4k}(x))\right\},
\label{eq:gauge}
\end{eqnarray}
where $i$, $j$, $k$ are for spatial directions and 
\begin{eqnarray}
P_{\mu\nu}(x)&=&1-\frac{1}{3} Re Tr \{ U_{\mu}(x)U_{\nu}(x+\hat{\mu})
\nonumber\\
&&\times U_{\mu}^\dagger(x+\hat{\nu})U_{\nu}^\dagger(x) \},
\\
R_{\mu\nu}(x)&=&1-\frac{1}{3} Re Tr \{ U_{\mu}(x)U_{\mu}(x+\hat{\mu})
U_{\nu}(x+2\hat{\mu})
\nonumber\\
&&\times
U_{\mu}^\dagger(x+\hat{\mu}+\hat{\nu})U_{\mu}^\dagger(x+\hat{\mu})
U_{\nu}^\dagger(x) \},
\end{eqnarray}
are plaquette and rectangular loop in the $\mu$-$\nu$ plane, respectively.
The improvement coefficients $c_i^{s/t}$ satisfy the normalization conditions, 
$c_0^s+8c_1^s=1$ and $c_0^t+4c_1^t+4c_2^t=1$. 
The bare gauge coupling equals $\beta=6/g^2$, 
and $\gamma_G$ represents the bare anisotropy.
We have three independent improvement parameters among $c_i^{s/t}$.

In principle, these improvement coefficients may have non-trivial 
$\xi$ dependences depending on 
the improvement conditions on the anisotropic lattice. 
In Ref.~\cite{ejiri02}, we have repeated the improvement procedure of 
Iwasaki on anisotropic lattices, 
and found that, for small anisotropies $\xi \approx 1$--4, 
the $\xi$-dependences in the improvement coefficients are weak, and a
sufficient improvement is achieved just by fixing the coefficients to 
Iwasaki's values for isotropic lattices, 
$c_1^s = c_1^t = c_2^t = -0.331$.
As explained in Sec.~\ref{sec:parameter}, we are interested in the case 
$\xi=2$.
Because $\xi$-dependences in the improvement coefficients require additional 
elaborations in numerical simulations, 
such as computation of $\xi$-derivative terms in the equation of state (EOS), 
we fix the improvement coefficients to their isotropic values
in the following.

\subsection{Clover quark action on anisotropic lattice}
\label{sec:Clover}

\begin{figure}
\resizebox{80mm}{!}{\includegraphics{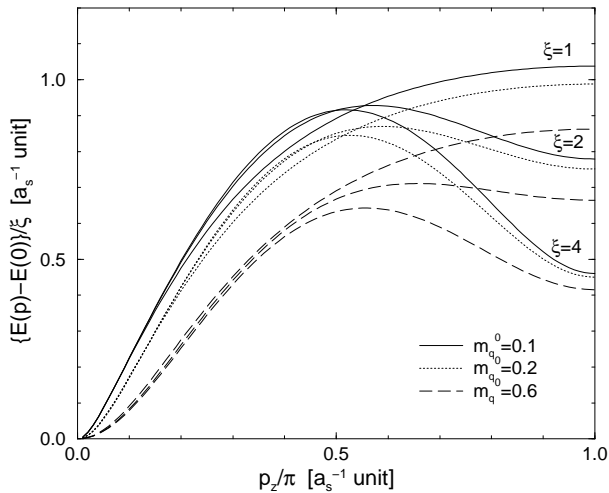}}
\caption{Dispersion relation of free quark on anisotropic lattices.
The energy $E(\vec{p})-E(0)$ normalized by $\xi$ is plotted as a 
function of spatial momentum at bare quark masses $m_q^0 =0.1$, 0.2 
and 0.6 (full, dotted and dashed curves) for anisotropies 
$\xi=1$, 2 and 4.
Larger $E(\vec{p})-E(0)/\xi$ at $p_z/\pi \sim 1$ correspond to 
smaller $\xi$.
}
\label{fig:free_disp}
\end{figure}

We employ clover-improved Wilson quarks \cite{clover}. 
On anisotropic lattices, the  action is given by
\begin{equation}
  S_F = \sum_{x,y} \bar{q}(x) K(x,y) q(y),
\label{eq:quark}
\end{equation}
\begin{eqnarray}
&& K(x,y) = \delta_{x,y} 
\nonumber\\
&& - \kappa_t \left\{ (1-\gamma_4)U_4(x)\delta_{x+\hat{4},y}
 +(1+\gamma_4)U^\dagger_4(x-\hat{4})\delta_{x-\hat{4},y} \right\}
\nonumber \\
&& - \kappa_s \sum_{i} 
   \left\{ (r-\gamma_i)U_i(x)\delta_{x+\hat{i},y} 
  + (r+\gamma_i)U^\dagger_i(x-\hat{i})\delta_{x-\hat{i},y} \right\}
\nonumber \\ 
&& - \kappa_s \left\{c_t \sum_i \sigma_{4i}F_{4i}(x)
               + r c_s \sum_{i>j}
                   \sigma_{ij}F_{ij}(x) \right\} \delta_{x,y}.
\end{eqnarray}
For the field strength $F_{\mu\nu}$, we use the standard clover-leaf
definition.
Following our previous studies at $\xi=1$, 
we apply a mean-field improvement for (\ref{eq:quark}), 
$U_i(x) \rightarrow U_i(x)/u_s$ and  $U_4(x) \rightarrow U_4(x)/u_t$, 
where $u_s$ and $u_t$ are mean links in the spatial and temporal directions.
For the mean links, we adopt the value estimated from 
plaquette in one-loop perturbation theory as in our previous studies. 
At $\xi=2$, we obtain
\begin{eqnarray}
 W_{11}(ss) &=& 1 - 1.154/\beta, 
\label{eq:wss}
\\
 W_{11}(st) &=& 1 - 0.560/\beta, 
\label{eq:wst}
\end{eqnarray} 
for spatial and temporal plaquettes.
Therefore, we set $u_{s}=(1-1.154/\beta)^{1/4}$ for $\xi=2$. 
For the temporal meanfield, we adopt $u_{t}=1$ because a naively calculated 
value $W_{11}(st)^{1/2}/W_{11}(ss)^{1/4}$
exceeds 1 for $\xi$ \gsim 1.6 at our values of $\beta$. 

Following Refs.~\cite{umeda01,taro00}, 
we set the spatial Wilson parameter to be $r=1/\xi$.
In this case, the quark dispersion relation in physical units preserves 
the 4-dimensional rotation symmetry at the tree level, and the tree-level 
improvement coefficients are free from the terms linear in $m_q$
\cite{harada01,aoki01,okamoto02}.
In a quenched study \cite{matsufuru01}, 
it was shown that the fermionic bare anisotropy parameter corresponding 
to a fixed $\xi$ is well fitted by a quadratic function of $m_q$ 
with this choice of $r$.
With dynamical quarks, however, terms linear in $m_q$ may appear 
through quark loop corrections \cite{aoki01}. 

The clover coefficients, $c_s$ and $c_t$, are functions of $r$.
With our choice $r=1/\xi$, $c_s$ and $c_t$ are unity at the tree level
and their mean-field improved values are given by
\begin{equation}
c_t=\frac{1}{u_s u_t^2}, \;\;\; c_s=\frac{1}{u_s^3}. 
\label{eq:csw}
\end{equation}

We define the bare anisotropy of the fermion field by 
\begin{equation}
\gamma_F = \frac{\kappa_t u_t}{\kappa_s u_s}.
\label{eq:gammaF}
\end{equation}
The bare quark mass in units of $a_s$ is given by
\begin{equation}
m_q^0 = \frac{1}{2\kappa_s u_s} - \gamma_F - 3r.
\end{equation}
For later convenience, we define $\kappa$ which satisfy 
the same relation with $m_q^0$ as in the isotropic case:
\begin{equation}
\frac{1}{\kappa} = 2(m_q^0 + 4)
= \frac{1}{\kappa_s u_s} - 2(\gamma_F+3r-4).
\label{eq:kappa}
\end{equation}
We perform chiral extrapolations in terms of $1/\kappa$.

The relation $r=1/\xi$ suggests that spatial doublers may appear 
at large $\xi$.
The free quark dispersion relation for our action is given by
\begin{equation}
 \cosh E(\vec{p}) =1+\frac{{\bar{\vec{p}}}^2 +
(\frac{m_q^0}{\gamma_F}+\frac{r}{2\gamma_F}{\hat{\vec{p}}}^2)^2}
{2(1+\frac{m_q^0}{\gamma_F}+\frac{r}{2\gamma_F}\hat{\vec{p}}^2)},
\end{equation}
where $\bar{p_i}=(\sin{p_i})/\gamma_F$, 
$\hat{p_i}=2\sin{(p_i/2)}$, and $E$ is in units of $a_t$ while 
$p_i$ is in units of $a_s$ \cite{matsufuru01}.
In Fig.~\ref{fig:free_disp}, we plot the energy $E(\vec{p})/\xi$ 
for $\xi=1$, 2 and 4 at $m_q^0 =0.1$, 0.2 and 0.6, where 
$\xi = \gamma_F$ in this approximation.
From this figure, we expect that doubler effects are weak 
at $\xi=2$ and $m_q = (1/\kappa - 1/\kappa_c)/2 = 0.07$--0.8 we study 
(see Sec.~\ref{sec:physics}).

\section{Simulation parameters}
\label{sec:parameter}

\begin{table*}
\caption{Simulation parameters. ``traj.'' is the number of trajectories 
used for measurements after 300 thermalization trajectories.
At $\beta=2.0$, besides the main simulation on the $10^3\times 30$ lattice, 
simulations using the same values of $\kappa$, $\gamma_F$, $\gamma_G$, and 
the trajectory length are done also on $8^3\times 24$ and $12^3\times 36$ 
lattices for a study of finite size effects.}
\begin{ruledtabular}
\begin{tabular}{ccccl}
 $\beta$ & size & $\kappa$ & traj. & $(\gamma_G,\gamma_F u_s)$ \\
\hline
 1.8 & $8^3\times 24$ & 0.10745 & 1700 & 
(1.70,0.90), (1.70,1.10), (1.70,1.20), (1.75,0.90), (1.75,1.10),
(1.75,1.20), \\
      &                &         &      & 
(1.85,0.90), (1.85,1.10)\\

      &                & 0.11162 & 1700 & 
(1.70,1.00), (1.70,1.20), (1.70,1.30), (1.75,1.00), (1.75,1.20), 
(1.75,1.30) \\

      &                & 0.11582 & 1700 & 
(1.70,1.15), (1.70,1.25), (1.70,1.30), (1.75,1.15), (1.75,1.25), 
(1.75,1.30) \\

      &                & 0.12115 & 1700 & 
(1.70,1.25), (1.70,1.35), (1.70,1.40), (1.75,1.25), (1.75,1.35), 
(1.75,1.40) \\

      &                & 0.12438 & 1700 & 
(1.70,1.20), (1.70,1.30), (1.70,1.40), (1.70,1.45), (1.75,1.30), 
(1.75,1.40), \\
      &                &         &      & 
(1.75,1.45), (1.80,1.20)\\

      &                & 0.12655 & 1700 & 
(1.70,1.35), (1.70,1.40), (1.70,1.45), (1.75,1.35), (1.75,1.40), 
(1.75,1.45) \\

\hline
 1.9 & $8^3\times 24$ & 0.1085 & 1000 & 
(1.80,1.00), (1.80,1.10), (1.80,1.20), (1.80,1.30), (1.85,1.00), 
(1.85,1.10), \\
      &                &         &      &
(1.85,1.20), (1.85,1.30), \\

      &                & 0.1137 & 1000 & 
(1.80,1.15), (1.80,1.25), (1.80,1.35), (1.85,1.15), (1.85,1.25), 
(1.85,1.35) \\

      &                & 0.1169 & 1000 & 
(1.75,1.20), (1.75,1.30), (1.75,1.40), (1.80,1.20), (1.80,1.30), 
(1.80,1.40), \\
      &                &         &     &
(1.85,1.20), (1.85,1.30)\\

      &                & 0.1212 & 1000 & 
(1.75,1.55), (1.80,1.25), (1.80,1.35), (1.80,1.45), (1.80,1.55),
(1.85,1.25), \\
      &                &         &      &
(1.85,1.35), (1.85,1.45) \\

      &                & 0.1245 & 1500 & 
(1.70,1.50), (1.70,1.60), (1.75,1.30), (1.80,1.40), (1.80,1.50), 
(1.85,1.30), \\
      &                &         &      & 
(1.85,1.60) \\

      &                & 0.1260 & 1500 & 
(1.75,1.40), (1.75,1.60), (1.80,1.50), (1.85,1.40), (1.85,1.60), 
(1.90,1.50) \\

\hline
 2.0 & $10^3\times 30$ & 0.1090 & 1000 & 
(1.80,1.25), (1.80,1.35), (1.80,1.45), (1.85,1.25), (1.85,1.35) \\

      & ($8^3\times 24$) & 0.1150 & 1000 & 
(1.80,1.45), (1.80,1.55), (1.85,1.35), (1.85,1.45), (1.85,1.55), 
(1.95,1.45)\\

      & ($12^3\times 36$) & 0.1180 & 1000 & 
(1.80,1.40), (1.80,1.50), (1.80,1.60), (1.85,1.50), (1.85,1.60) \\

      &                 & 0.1210 & 1000 & 
(1.80,1.45), (1.80,1.55), (1.80,1.65), (1.85,1.45), (1.85,1.55), 
(1.95,1.45)\\

      &                 & 0.1244 & 1500 & 
(1.70,1.60), (1.80,1.50), (1.80,1.60), (1.80,1.70), (1.85,1.55), 
(1.85,1.60), \\
      &                 &        &      & 
(1.90,1.55), (1.90,1.60), (2.00,1.50)\\

      &                 & 0.1252 & 1500 & 
(1.75,1.60), (1.75,1.65), (1.80,1.60), (1.85,1.55), (1.85,1.65) \\

\hline
 2.1 & $12^3\times 36$ & 0.1100 & 1000 & 
(1.80,1.35), (1.80,1.55), (1.90,1.45), (1.95,1.35) \\

      &                & 0.1150 & 1000 & 
(1.80,1.50), (1.80,1.60), (1.90,1.45), (1.90,1.55), (1.90,1.65) \\

      &                & 0.1200 & 1000 & 
(1.80,1.65), (1.85,1.55), (1.90,1.75), (1.95,1.50), (1.95,1.60) \\

      &                & 0.1225 & 1500 & 
(1.80,1.60), (1.80,1.70), (1.80,1.80), (1.90,1.60), (1.90,1.70), 
(1.90,1.80) \\

      &                & 0.1245 & 1500 & 
(1.80,1.60), (1.80,1.80), (1.85,1.70), (1.90,1.60), (1.90,1.70) \\

\end{tabular}
\label{tab:param}
\end{ruledtabular}
\end{table*}

In this paper, 
we focus on the case of the renormalized anisotropy $\xi=2$.
We have shown for finite-temperature pure SU(3) gauge theory 
\cite{namekawa01} that  
this choice of $\xi$ is optimal to reduce scaling violations
in equation of state both in the high temperature limit and 
at finite $\beta$; the latter is confirmed by a Monte Carlo simulation. 
It is straightforward to analyze the high temperature limit for full 
QCD.  We have found that $\xi=2$ is optimal also with two 
flavors of dynamical quarks and improved glue.

We perform simulations at $\beta=1.8$, 1.9, 2.0, and 2.1 on
$8^3\times24$, $8^3\times24$, $10^3\times30$ and $12^3\times36$ 
lattices, respectively. 
The lattice spacing is in the range $a_s \approx 0.28$--0.16 fm, 
and hence the spatial lattice size is fixed to be about 2 fm. 
See Sec.~\ref{sec:physics} for details of the scale determination.
At each $\beta$, six values of $\kappa$, corresponding to 
$m_{PS}/m_{V}\approx 0.6$, 0.7, 0.8, 0.85, 0.9, and 0.92, are simulated. 
To study lattice volume effects, we also perform additional simulations 
on $8^3\times24$ and $12^3\times36$ lattices at $\beta=2.0$. 
Our simulation parameters are summarized in Table~\ref{tab:param}.

We generate gauge configurations by the HMC algorithm 
with an even-odd preconditioned BiCGStab quark solver \cite{cppacs02}.
The molecular dynamics time step $dt$ is adjusted to achieve 
an acceptance rate of about 70--80\%.
Measurements are performed at every 5 trajectories 
over 1000--1700 trajectories after 300 thermalization trajectories, 
where the length of one trajectory is set to unity. 
Statistical errors of the observables are estimated by the jackknife 
method at each $\beta$ and $\kappa$ with bins of 50 trajectories.

\section{Calibration procedure}
\label{sec:calib_proc}

At each $\beta$ and $\kappa$,
we have to tune the bare anisotropy parameters $\gamma_F$ and $\gamma_G$ 
such that the renormalized anisotropies $\xi_F$ and $\xi_G$ for 
fermionic and gluonic observables coincide with each other:
\begin{equation}
 \xi_F(\gamma_F,\gamma_G;\beta,\kappa) = 
 \xi_G(\gamma_F,\gamma_G;\beta,\kappa) = \xi .
\label{eq:xi2}
\end{equation}
As discussed in the previous section, we study the case $\xi=2$.
For this purpose, we measure $\xi_F$ and $\xi_G$ at several values of 
$(\gamma_F,\gamma_G)$ at fixed $\kappa$ and $\beta$,
and determine the point where Eq.~(\ref{eq:xi2}) is satisfied 
by an interpolation in $\gamma_F$ and $\gamma_G$. 
Let us denote the resulting values of $\gamma_F$ and $\gamma_G$ for 
$\xi=2$ as $\gamma_F^*(\beta,\kappa)$ and $\gamma_G^*(\beta,\kappa)$.
Finally, we parametrize $\gamma_F^*$ and $\gamma_G^*$ as functions of 
$\beta$ and $\kappa$ for use in future studies of heavy quark systems 
and thermodynamics of QCD.

We measure $\xi_G$ by Klassen's method \cite{klassen98}: 
\begin{equation}
 R_s(x,y) = R_t(x,\xi_Gy),
\end{equation}
where 
\begin{eqnarray}
 R_s(x,y)&=&\frac{W_{ss}(x,y)}{W_{ss}(x+1,y)}, 
\\
 R_t(x,t)&=&\frac{W_{st}(x,t)}{W_{st}(x+1,t)},
\end{eqnarray}
are the ratios of spatial-spatial and spatial-temporal Wilson loops, 
$W_{ss}(x,y)$ and $W_{st}(x,t)$, respectively.
We determine $\xi_G$ by minimizing
\begin{equation}
L(\xi_G) = \sum_{x,y}\frac{(R_s(x,y)-R_t(x,\xi_G y))^2}
{(\Delta R_s)^2+(\Delta R_t)^2},
\label{eq:indicator}
\end{equation}
with $\Delta R_s$ and $\Delta R_t$ the statistical errors of $R_s$ and $R_t$.
To avoid short range lattice artifacts, $x$ and $y$ should not be too small.
Practical range of $x$ and $y$ will be discussed later.

For $\xi_F$ we use the relativistic dispersion relation of mesons: 
\begin{equation}
 E(\vec{p})^2=m^2+\frac{\vec{p}^2}{\xi_F^2}+O(\vec{p}^4),
\label{eq:disp}
\end{equation}
where $E$ and $m$ are the energy and mass in units of $a_t$, 
and $\vec{p}=2\pi\vec{n}/L_s$, with $L_s$ the spatial lattice size,
is the spatial momentum in units of $a_s$.
We evaluate $E$ and $m$ from a cosh fit of the meson two-point correlation
function,
\begin{eqnarray}
 C(\vec{p},t)&=&\sum_{\vec{x}}\langle O(\vec{x},t)
O^\dagger(\vec{0},0) e^{i\vec{p}\vec{x}} \rangle \\
O(\vec{x},t)&=&\sum_{\vec{y}\vec{z}}\phi(\vec{y})\phi'(\vec{z})
\bar{q}(\vec{x}+\vec{y},t)\Gamma q(\vec{x}+\vec{z},t)
\end{eqnarray}
In this paper, we study pseudoscalar (PS) and vector (V) mesons 
consisting of sea quarks only:
$\Gamma = \gamma_5$ for PS and $\Gamma = \gamma_i$ for V.
Quark fields are smeared by 
a function $\phi(\vec{x})$ to enhance ground state 
signals at short distances.
For the ``smeared'' quark field, 
we adopt an exponential smearing function of form
\begin{eqnarray}
\phi(\vec{x}) = a \exp{(-p|\vec{x}|)} \;\; {\rm for} \; \vec{x}\neq0, 
\;\;\;
\phi(\vec{0}) = 1,
\end{eqnarray}
where the coefficients $a$ and $p$ are adopted from a previous study 
\cite{cppacs02}.
The ``point'' quark field corresponds 
to $\phi(\vec{x})=\delta_{\vec{x},\vec{0}}$.
In our calculation of meson two-point function, 
the sink operator is always the point-point type, 
while, for the source operator, we study point-point, point-smeared and 
smeared-smeared cases.
We find that the smear-smear source operator leads to the earliest plateau 
with small errors. 
Therefore, we adopt the smear-smear source operator.

In principle, we may adopt different observables to define 
the renormalized anisotropies. 
Away from the continuum limit, different choices will lead to 
$O(a)$ differences in the calibration results. 
To study this problem, we compare the calibration results using $\xi_F$ 
from PS and V meson dispersion relations. 
We denote these results for the calibrated bare anisotropies 
as $(\gamma_F^*(PS),\gamma_G^*(PS))$ and $(\gamma_F^*(V),\gamma_G^*(V))$,
respectively.
In Sec.~\ref{sec:bare}, we show that they tend to converge together 	 
toward the continuum limit.
In future applications of the present work, 
different sets of $(\gamma_F^*,\gamma_G^*)$ will be useful 
for estimating systematic errors due to the continuum extrapolation,
in complicated physical observables, such as the equation of state.

In a previous study of quenched QCD \cite{taro00}, 
the ratio of temporal and screening masses of the PS meson 
was used to determine $\xi_F$. 
We study the difference between our procedure and the mass ratio method 
in Appendix A. 
We find that both methods give consistent values of $\xi_F$ when 
the quarks are not too heavy 
[$m_{PS}/m_{V}$ \lsim 0.75 (0.8) at $\beta$ \gsim 2.0 (2.1)].

\section{Calibration results}
\label{sec:calib_result}

\subsection{$\xi_G$ from matching of Wilson loop ratios}
\label{sec:matching}

\begin{figure}
\resizebox{73mm}{!}{\includegraphics{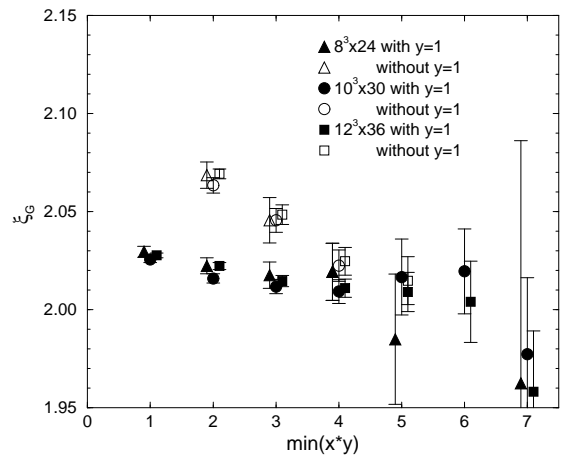}}
\caption{A typical for a determination of $\xi_G$
at $\beta=2.0$, $\kappa=0.1244$ and 
$(\gamma_G,\gamma_F u_s)=(1.85,1.60)$.
The $\xi_G$ shows the minimizing point of $L(\xi_G)$ defined by 
Eq.\ref{eq:indicator}. 
Different symbols represent the results obtained on 
$8^3\times 24$, $10^3\times 30$ and $12^2\times 36$ lattices.
For filled/open symbols, $L(\xi_G)$ is evaluated with/without 
the $y=1$ data.}
\label{fig:indicator}
\end{figure}

We determine the renormalized gauge anisotropy $\xi_G$ by minimizing
the function $L(\xi_G)$ defined by Eq.~(\ref{eq:indicator}). 
We interpolate $R_t(x,t)$ by a cubic spline in terms of $t$.
To remove short range lattice artifacts,
we evaluate $L(\xi_G)$ with $x$ and $y$ which satisfy 
$x\times y \ge M$ and examine the $M$ dependence. 
The upper limit on $x$ and $y$ is set by requiring that the statistical 
error does not exceed the central value for the Wilson loop ratio. 
Varying the upper limit hardly changes results for $\xi_G$. 
Filled symbols in Fig.~\ref{fig:indicator} show typical results of 
$\xi_G$ as a function of $M=\min(x\times y)$. 
We find that, at this simulation point,
$\xi_G$ is reasonably stable when $x\times y$ is larger than about 4 .

Since the condition $x\times y \ge M$ does not exclude small $x$ or $y$, 
which can be an additional origin of short 
distance effects, we study whether $\xi_G$ are affected 
by small values of $x$ or $y$ by removing them. 
The results of $\xi_G$ using $L(\xi_G)$ without data at $y=1$
are plotted with open symbols in Fig.~\ref{fig:indicator}.
We find that, although clear deviations from the filled symbols are 
observed at small $M$, the effects of $y=1$ data are  
within 1\% at $M \gsim 4$ where $\xi_G$ becomes stable.
Therefore, placing a condition on $\min(x\times y)$ is sufficient to obtain a 
stable value for $\xi_G$.
Similar results are obtained at other simulation points.

Results obtained on different lattice volumes 
($8^3\times 24$, $10^3\times 30$ and $12^2\times 36$)
are also shown in Fig.~\ref{fig:indicator}.
With our lattices, no finite volume effects are visible in 
the values of $\xi_G$.

From these studies, we adopt 
$\min(x\times y) = 3$, 3, 4, and 5 at $\beta = 1.8$, 1.9, 2.0 
and 2.1, respectively, in the subsequent analyses.

\subsection{$\xi_F$ from meson dispersion relations}
\label{sec:dispersion}

\begin{figure}
\resizebox{73mm}{!}{\includegraphics{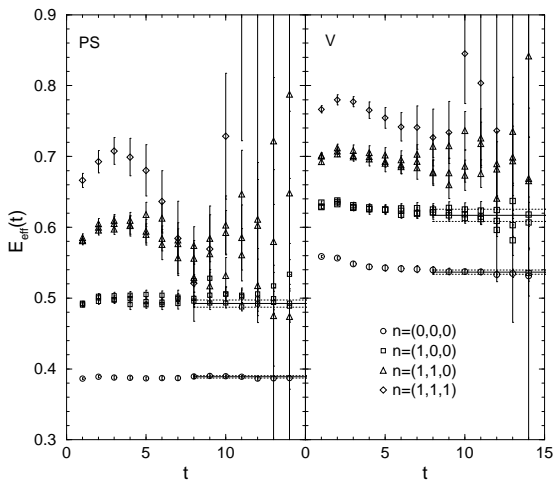}}
\caption{Effective mass of meson states with various momenta obtained  
at $\beta=2.0$, $\kappa=0.1244$, $(\gamma_G,\gamma_F u_s)=(1.85,1.60)$
on the $10^3\times30$ lattice. 
The left and right panels are the results for pseudoscalar and vector 
mesons. 
}
\label{fig:effective}
\end{figure}

\begin{figure}
\resizebox{73mm}{!}{\includegraphics{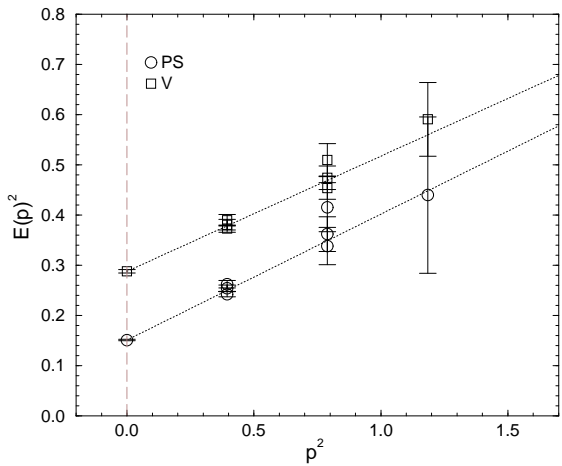}}
\caption{Sample results for the dispersion relation of pseudoscalar 
and vector mesons at $\beta=2.0$, $\kappa=0.1244$ and 
$(\gamma_G,\gamma_F u_s)=(1.85,1.60)$.
Dotted lines show fit results from $\vec{n}=(0,0,0)$ and $(1,0,0)$.}
\label{fig:disp}
\end{figure}

\begin{figure}
\resizebox{73mm}{!}{\includegraphics{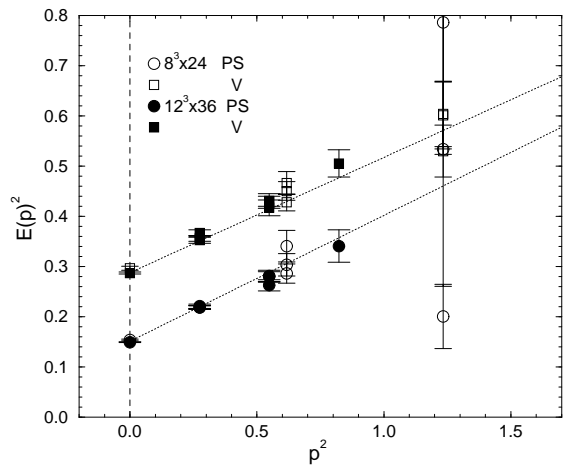}}
\caption{ Volume dependence of a mesonic dispersion relations
at $\beta=2.0$, $\kappa=0.1244$ and 
$(\gamma_G,\gamma_F u_s)=(1.85,1.60)$.
Filled and open symbols show the results on $12^3\times 36$ and 
$8^3\times 24$ lattices with PS and V channels. 
Other condition is same as Fig.\ref{fig:disp} and fit results 
on $10^3\times30$, which are the same as Fig.\ref{fig:disp}, are
shown with dotted lines.
}
\label{fig:v_dep}
\end{figure}

We determine the renormalized quark anisotropy $\xi_F$ from the meson 
dispersion relation.
We calculate the meson energy $E(\vec{p})$ at the spatial momenta 
$\vec{p}=2\pi \vec{n}/L_s$
with $\vec{n} = (0,0,0)$, (1,0,0), (1,1,0), (1,1,1), and their 
permutations.
In Fig.~\ref{fig:effective} we plot typical data for effective energy 
defined by
\begin{equation}
 \frac{C(\vec{p},t)}{C(\vec{p},t+1)}
=\frac{\cosh{[E_{\rm eff}(\vec{p},t)(N_t/2-t)]}}
{\cosh{[E_{\rm eff}(\vec{p},t)(N_t/2-t-1)]}}.
\end{equation}
obtained from the smear-smear correlators.  
Typical results for the energy $E(\vec{p})$ are shown in 
Fig.~\ref{fig:disp}.

Using data at $\vec{n}= (0,0,0)$, (1,0,0) and its permutations, 
we fit $E(\vec{p})$ with the leading formula 
$E(\vec{p})^2=m^2+\vec{p}^2 / \xi_F^2$ to determine $\xi_F$.
The fits are shown by dotted lines in Fig.~\ref{fig:disp}. 
In Fig.~\ref{fig:v_dep}, data obtained on $8^3\times 24$ and $12^2\times 36$
lattices are compared with the fit results on the $10^3\times30$ lattice.
We find that the data are well explained by the fit results. 
Indeed, the slope obtained for the three lattice sizes are consistent: 
$\xi_F=2.044(59)$, 2.020(44) and 2.012(39) 
on $8^3\times 24$, $10^3\times 30$ and $12^2\times 36$ lattices,
respectively. 
This confirms that the spatial lattice size \gsim 1.6 fm 
is sufficiently large to suppress finite volume effects in $\xi_F$ 
in the range of quark masses we study.

\subsection{Bare anisotropies at $\xi=2$ 
($\gamma_G^*$ and $\gamma_F^*$)}
\label{sec:bare}

\begin{figure}
\resizebox{82mm}{!}{\includegraphics{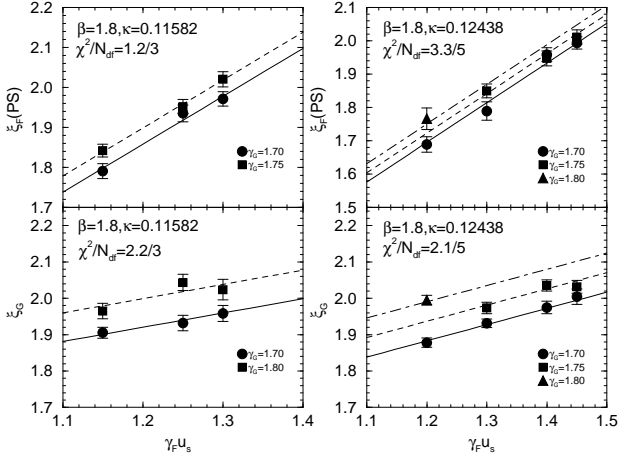}}
\caption{$\xi_G$ and $\xi_F$ as functions of $(\gamma_G,\gamma_F u_s)$
at $\beta=1.8$ for the third and fifth heaviest $\kappa$. 
For $\xi_F$, results from the pseudoscalar dispersion relation are shown.
Lines represent the results of the fits (\protect\ref{eq:xiFlinear})
and (\protect\ref{eq:xiGlinear}).
}
\label{fig:fit2d_b18}
\end{figure}

\begin{figure}
\resizebox{82mm}{!}{\includegraphics{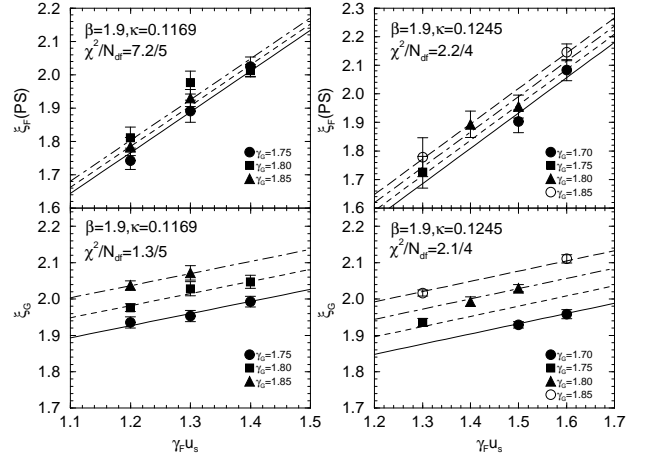}}
\caption{The same as Fig.~\protect\ref{fig:fit2d_b18} but at $\beta=1.9$. 
}
\label{fig:fit2d_b19}
\end{figure}

\begin{figure}
\resizebox{82mm}{!}{\includegraphics{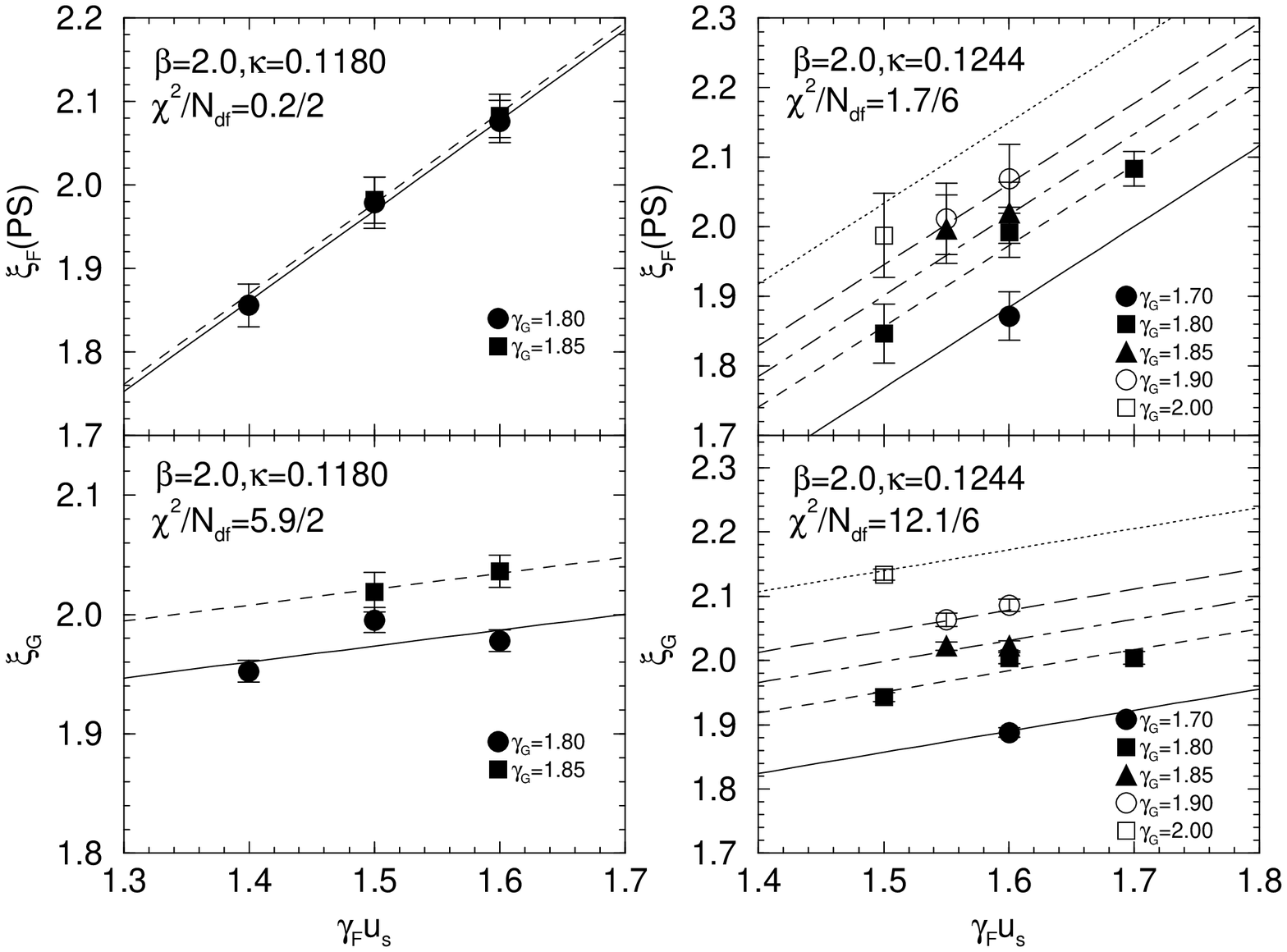}}
\caption{The same as Fig.~\protect\ref{fig:fit2d_b18} but at $\beta=2.0$. 
}
\label{fig:fit2d_b20}
\end{figure}

\begin{figure}
\resizebox{82mm}{!}{\includegraphics{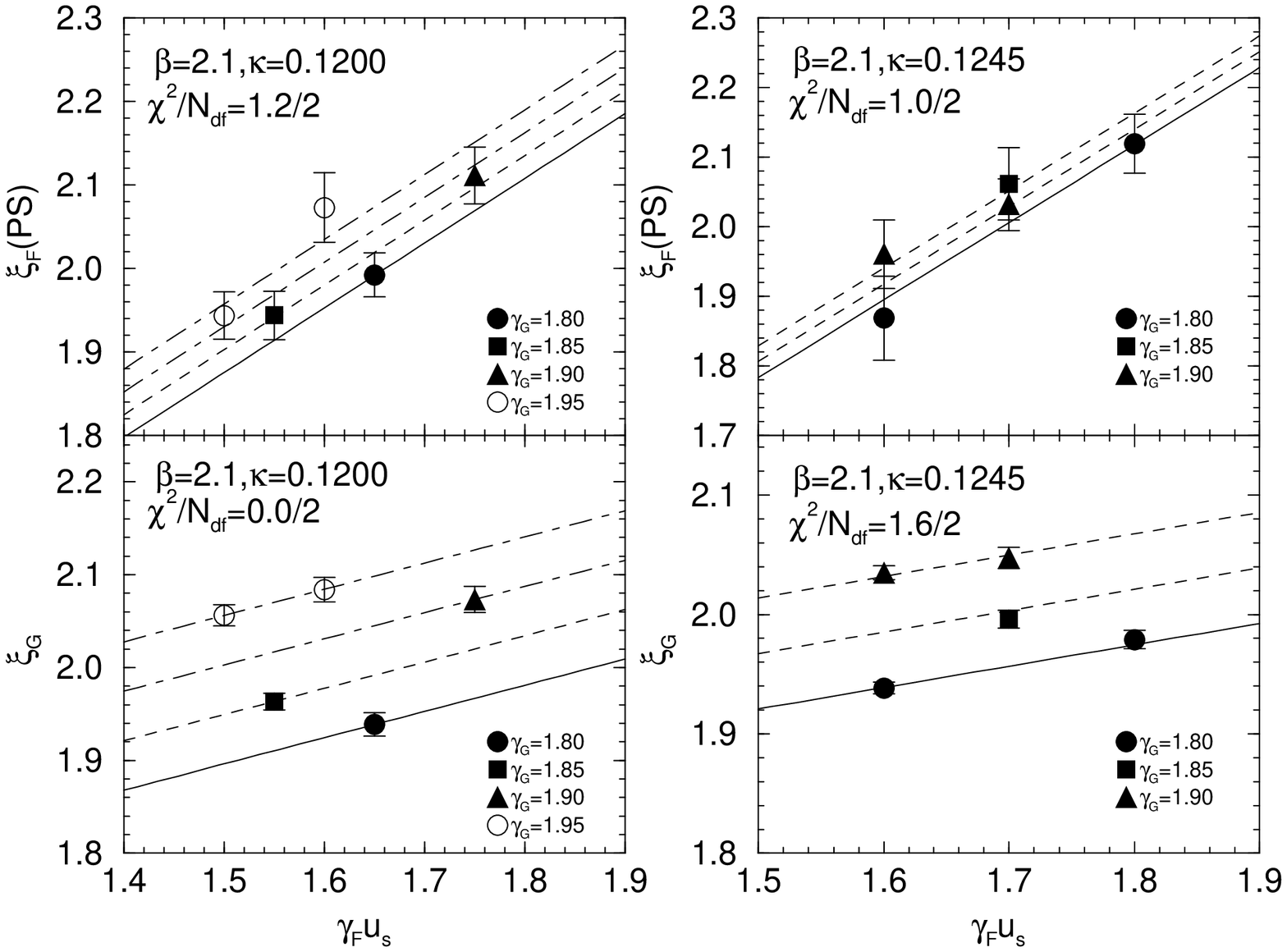}}
\caption{The same as Fig.~\protect\ref{fig:fit2d_b18} but at $\beta=2.1$. 
}
\label{fig:fit2d_b21}
\end{figure}

\begin{figure}
\resizebox{82mm}{!}{\includegraphics{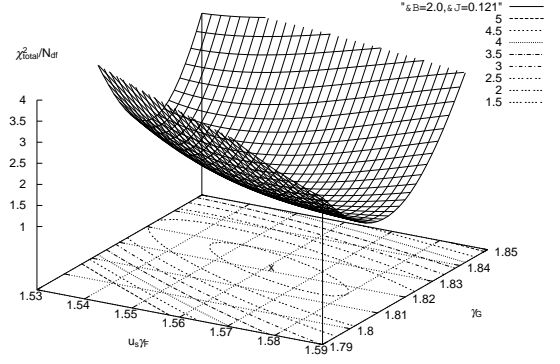}}
\caption{$\chi^2_{total}/N_{df}$ for $\xi=2$ fits at $\beta=2.0$ and 
$\kappa=0.121$. Results using $\xi_F$(PS) for the fermionic anisotropy 
are shown.
The minimum of $\chi^2_{total}/N_{df}$ is 1.348 at 
$(u_s\gamma_F,\gamma_G) = (u_s\gamma_F^*,\gamma_G^*) = (1.5605, 1.8225)$.
Curves on the base plane is a contour map of $\chi^2_{total}/N_{df}$. 
The minimum point is marked by ``x'' on the contour map. 
[Note that $N_{df}$ is larger than that for the fits 
(\protect\ref{eq:xiFlinear}) and (\protect\ref{eq:xiGlinear}) 
summarized in Table~\protect\ref{tab:calib} because $a_F$ and $a_G$ are 
not free in this calculation.]
}
\label{fig:chisq_total}
\end{figure}

\begin{figure}
\resizebox{73mm}{!}{\includegraphics{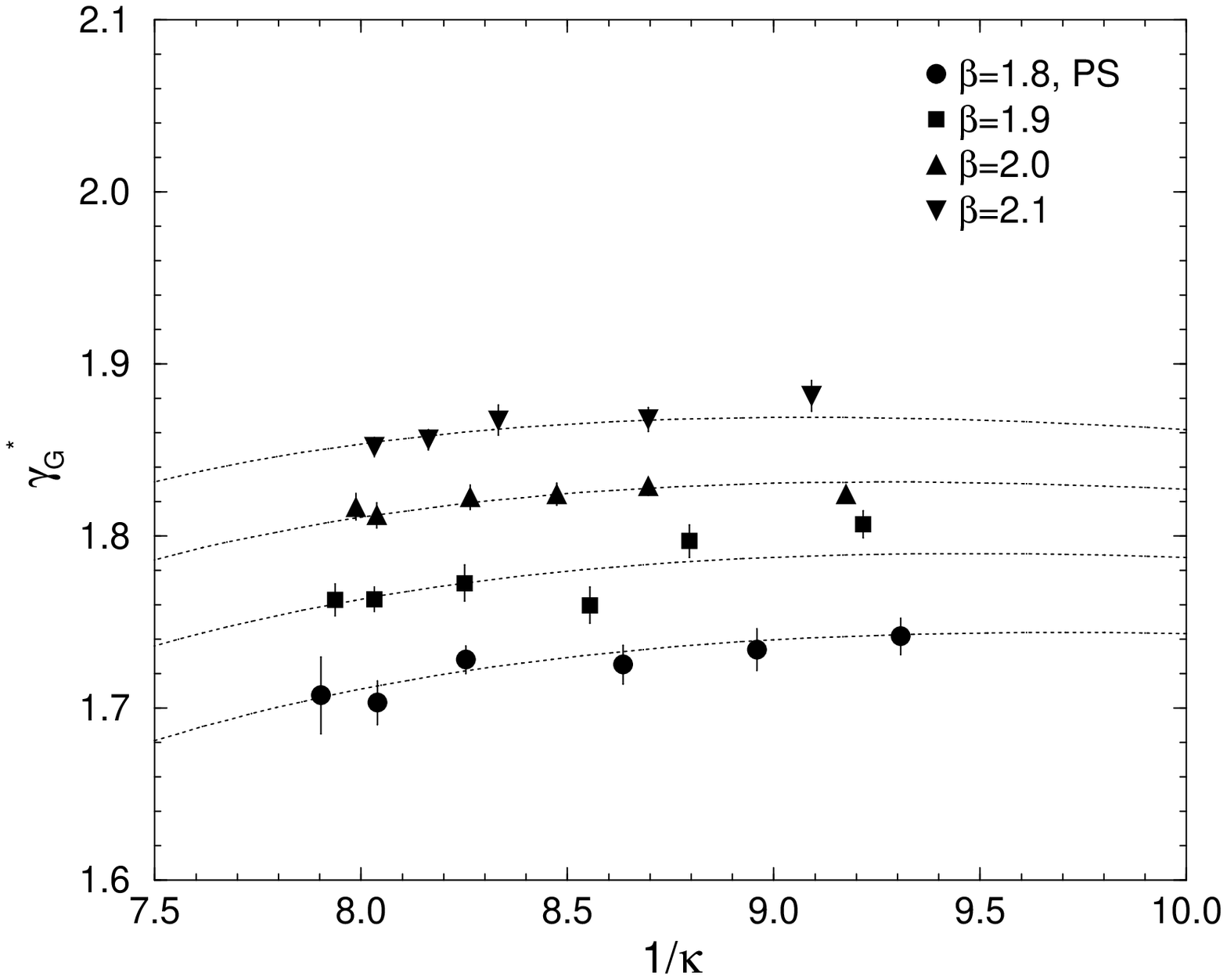}}
\resizebox{73mm}{!}{\includegraphics{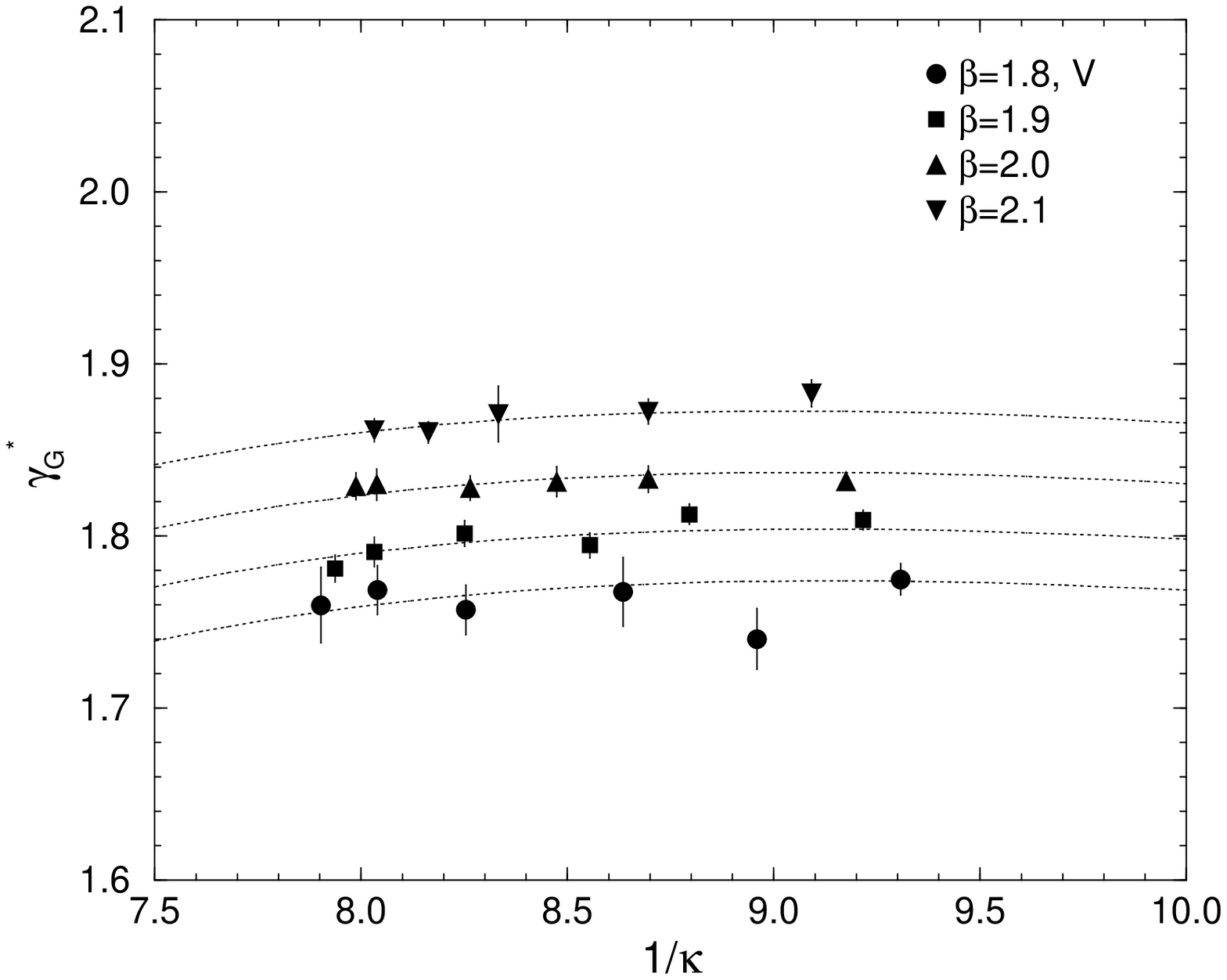}}
\caption{ $\gamma_G^*$(PS) and $\gamma_G^*$(V) corresponding to $\xi=2$.
Curves are the results of the global fit (\ref{eq:gammaG_form})
with parameters (\ref{eq:gammaGPS}) and (\ref{eq:gammaGV}).
}
\label{fig:gamma_g}
\end{figure}

\begin{figure}
\resizebox{73mm}{!}{\includegraphics{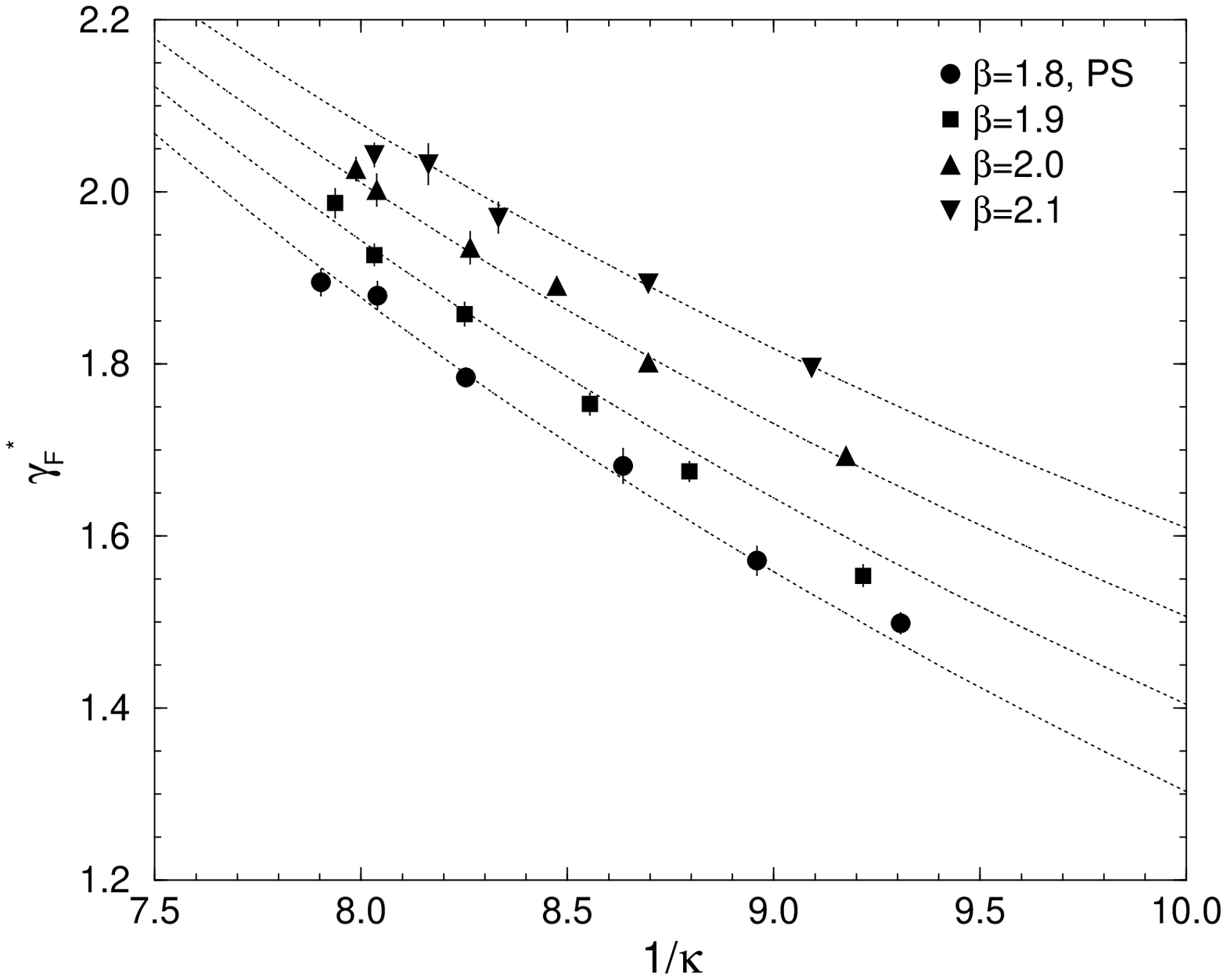}}
\resizebox{73mm}{!}{\includegraphics{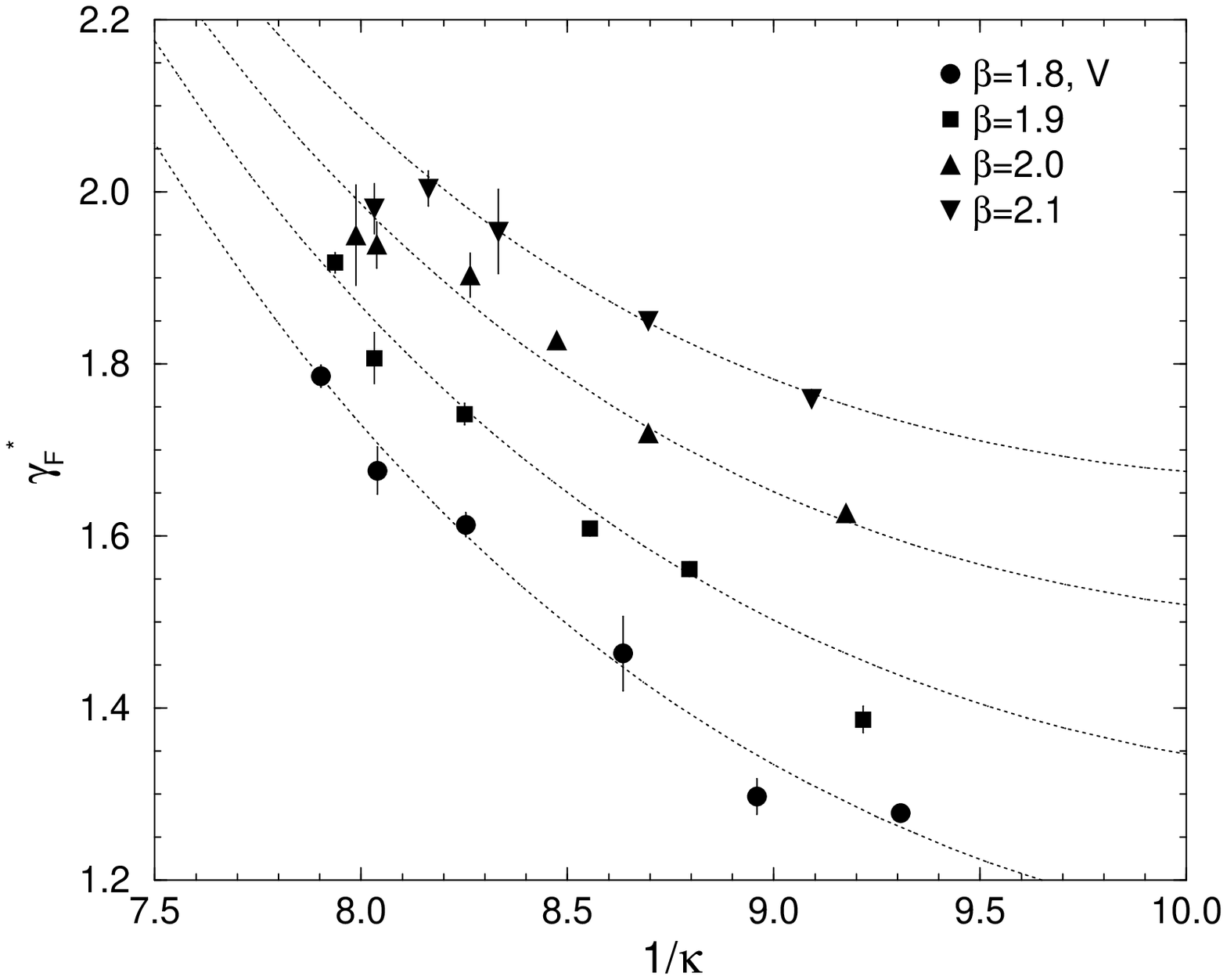}}
\caption{ $\gamma_F^*$(PS) and $\gamma_F^*$(V) corresponding to $\xi=2$.
Curves are the results of the global fit (\ref{eq:gammaF_form})
with parameters (\ref{eq:gammaFPS}) and (\ref{eq:gammaFV}).
}
\label{fig:gamma_f}
\end{figure}

\begin{figure}
\resizebox{73mm}{!}{\includegraphics{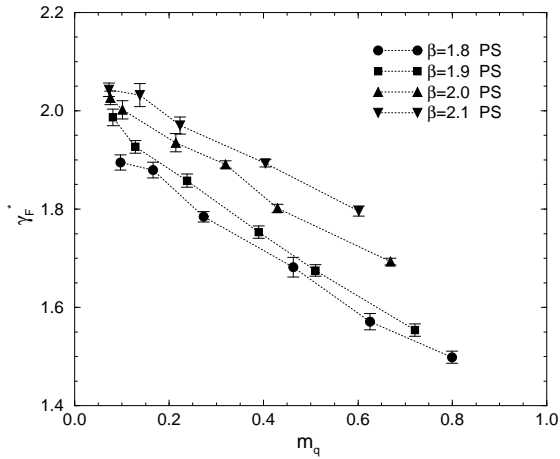}}
\caption{ $\gamma_F^*$(PS) vs quark mass at various $\beta$. 
Lines are guide of the eyes.
}
\label{fig:gamma_f_qmass}
\end{figure}

\begin{table*}
\caption{Bare anisotropy parameters calibrated to $\xi=2$.}
\begin{ruledtabular}
\begin{tabular}{cccccccccc}
 $\beta$ & $\kappa$
 & $\gamma_G^*$(PS) & $\chi^2/N_{df}$ 
 & $\gamma_F^*$(PS) & $\chi^2/N_{df}$
 & $\gamma_G^*$(V)  & $\chi^2/N_{df}$  
 & $\gamma_F^*$(V)  & $\chi^2/N_{df}$\\
\hline
 1.8 &  0.10745 & 
 1.742(11) & 12.4/ 5 & 1.499(12) &  6.7/ 5 & 1.7748(93) & 12.4/ 5 & 1.2783(82) &  3.1/ 5\\
      &  0.11162 & 
 1.734(12) &  2.7/ 3 & 1.571(17) &  2.6/ 3 & 1.740(18) &  2.7/ 3 & 1.298(21) &  2.0/ 3\\
      &  0.11582 & 
 1.725(11) &  2.2/ 3 & 1.682(20) &  1.2/ 3 & 1.768(20) &  2.2/ 3 & 1.464(43) &  2.4/ 3\\
      &  0.12115 & 
 1.7282(81) &  0.9/ 3 & 1.784(11) &  1.4/ 3 & 1.757(15) &  0.9/ 3 & 1.613(14) &  7.8/ 3\\
      &  0.12438 & 
 1.703(13) &  2.1/ 5 & 1.879(16) &  3.3/ 5 & 1.769(14) &  2.1/ 5 & 1.676(28) &  2.5/ 5\\
      &  0.12655 & 
 1.708(22) &  1.1/ 3 & 1.895(15) &  4.9/ 3 & 1.760(22) &  1.1/ 3 & 1.786(13) &  1.4/ 3\\
\hline
 1.9 &  0.10850 & 
 1.8068(80) &  1.6/ 5 & 1.554(12) &  5.1/ 5 & 1.8093(56) &  1.6/ 5 & 1.387(15) &  5.8/ 5\\
      &  0.11370 & 
 1.7971(94) &  2.9/ 3 & 1.675(12) &  1.3/ 3 & 1.8126(60) &  2.9/ 3 & 1.5615(84) &  1.6/ 3\\
      &  0.11690 & 
 1.760(11) &  1.3/ 5 & 1.753(13) &  7.2/ 5 & 1.7945(74) &  1.3/ 5 & 1.6085(82) &  3.8/ 5\\
      &  0.12120 & 
 1.773(10) &  4.0/ 5 & 1.858(13) &  1.6/ 5 & 1.8013(75) &  4.0/ 5 & 1.742(13) &  7.3/ 5\\
      &  0.12450 & 
 1.7631(72) &  2.1/ 4 & 1.927(13) &  2.2/ 4 & 1.7908(85) &  2.1/ 4 & 1.807(30) &  1.5/ 4\\
      &  0.12600 & 
 1.7629(93) &  2.8/ 3 & 1.987(17) &  1.8/ 3 & 1.7811(77) &  2.8/ 3 & 1.918(12) &  4.9/ 3\\
\hline
 2.0 &  0.10900 & 
 1.8243(40) &  2.9/ 2 & 1.6931(70) &  4.1/ 2 & 1.8319(38) &  2.9/ 2 & 1.6266(95) &  1.3/ 2\\
      &  0.11500 & 
 1.8288(52) &  1.5/ 3 & 1.8015(85) &  0.9/ 3 & 1.8331(76) &  1.5/ 3 & 1.7190(98) &  6.7/ 3\\
      &  0.11800 & 
 1.8243(65) &  5.9/ 2 & 1.8907(74) &  0.2/ 2 & 1.8315(88) &  5.9/ 2 & 1.8268(83) &  0.1/ 2\\
      &  0.12100 & 
 1.8225(73) &  7.5/ 3 & 1.935(19) &  3.3/ 3 & 1.8278(71) &  7.5/ 3 & 1.903(25) &  2.5/ 3\\
      &  0.12440 & 
 1.8120(73) & 12.1/ 6 & 2.002(19) &  1.7/ 6 & 1.8299(92) & 12.1/ 6 & 1.938(27) &  3.1/ 6\\
      &  0.12520 & 
 1.8169(76) &  2.2/ 2 & 2.026(14) &  0.4/ 2 & 1.8289(78) &  2.2/ 2 & 1.949(58) &  5.5/ 2\\
\hline
 2.1 &  0.11000 & 
 1.8814(88) &  2.7/ 1 & 1.796(10) &  1.4/ 1 & 1.8827(79) &  2.7/ 1 & 1.760(11) &  0.4/ 1\\
      &  0.11500 & 
 1.8678(70) &  4.5/ 2 & 1.8932(74) & 17.5/ 2 & 1.8722(73) &  4.5/ 2 & 1.8501(85) & 19.4/ 2\\
      &  0.12000 & 
 1.8673(87) &  0.0/ 2 & 1.970(18) &  1.2/ 2 & 1.871(16) &  0.0/ 2 & 1.954(49) &  0.6/ 2\\
      &  0.12250 & 
 1.8559(58) &  2.6/ 3 & 2.032(23) &  3.6/ 3 & 1.8603(62) &  2.6/ 3 & 2.004(20) &  6.0/ 3\\
      &  0.12450 & 
 1.8517(55) &  1.6/ 2 & 2.043(14) &  1.0/ 2 & 1.8615(68) &  1.6/ 2 & 1.980(29) &  4.8/ 2\\
\end{tabular}
\end{ruledtabular}
\label{tab:calib}
\end{table*}

Figures~\ref{fig:fit2d_b18}--\ref{fig:fit2d_b21} show typical results for 
bare anisotropies at each $\beta$, obtained at the third and fifth heaviest
quark masses ($m_{PS}/m_V \sim 0.85$ and 0.70). 
We find that, for the range of parameters we study, we can fit the 
data assuming an ansatz linear in $\gamma_F$ and $\gamma_G$:
\begin{eqnarray}
\xi_F &=& a_F + b_F\gamma_F + c_F\gamma_G,
\label{eq:xiFlinear}\\
\xi_G &=& a_G + b_G\gamma_F + c_G\gamma_G.
\label{eq:xiGlinear}
\end{eqnarray}
Results of the least $\chi^2$ fits are also shown in 
Figs.~\ref{fig:fit2d_b18}--\ref{fig:fit2d_b21}.

From the condition
$\xi_F(\gamma_F^*,\gamma_G^*)=\xi_G(\gamma_F^*,\gamma_G^*)=2$, we
obtain $\gamma_F^*$ and $\gamma_G^*$ for $\xi=2$ as functions 
of the coefficients, $a_F,\cdots,c_G$. 
We determine their errors using the error propagation formula 
where the errors for $a_F,\cdots,c_G$ are estimated from the error
matrix of the least $\chi^2$ fits for $\xi_F$ and $\xi_G$. 
The results are summarized in Table~\ref{tab:calib}. 

To confirm the magnitude of errors, we study 
$\chi^2_{total}/N_{df} \equiv (\chi^2_{F;2}+\chi^2_{G;2})/2N_{df}$, 
as a function of $\gamma_F$ and $\gamma_G$, where
$\chi^2_{F/G;2}$ is the $\chi^2$ value for a fit of 
$\xi_{F/G}(\gamma'_F,\gamma'_G)$ data to 
$\xi_{F/G} = 2 + b (\gamma'_F - \gamma_F) + c (\gamma'_G - \gamma_G)$
for given values of $(\gamma_F,\gamma_G)$.
This quantity measures to which extent $\xi=2$ is achieved by Wilson loops
and the meson correlation function at $(\gamma_F,\gamma_G)$.
The minimum of $\chi^2_{total}/N_{df}$ locates at $(\gamma_F^*,\gamma_G^*)$. 
A typical result is plotted in Fig.~\ref{fig:chisq_total}.
We find that errors estimated from a unit increase of $\chi^2_{total}/N_{df}$ 
are consistent with those listed in Table~\ref{tab:calib}.

In later applications, it will be convenient to parametrize $\gamma_F^*$
and $\gamma_G^*$ as functions of $\beta$ and $\kappa$. 
Figures~\ref{fig:gamma_g} and \ref{fig:gamma_f} show the parameter 
dependence of $\gamma_F^*$ and $\gamma_G^*$. 
We adopt the general quadratic ansatz in $\beta$ and $\kappa$,
\begin{eqnarray}
\gamma_F^* &=& A_F + B_F\beta' + C_F\beta'^2 
\nonumber \\
&& + D_F\beta'\kappa' + E_F\kappa' + F_F\kappa'^2,
\label{eq:gammaF_form}\\
\gamma_G^* &=& A_G + B_G\beta' + C_G\beta'^2 
\nonumber \\
&& + D_G\beta'\kappa' + E_G\kappa' + F_G\kappa'^2,
\label{eq:gammaG_form}
\end{eqnarray}
where $\beta' = \beta - 2.0$ and $\kappa' = \kappa - 0.12$.
For $\gamma_F^*$(PS) and $\gamma_G^*$(PS), we find
\begin{eqnarray}
A_F &=& 1.9097(41), \; B_F = 0.746(38), 
\nonumber \\
C_F &=& 0.04(25), \; D_F = -13.9(4.1), 
\nonumber \\
E_F &=& 20.14(75), \; F_F = -1.(83.), 
\label{eq:gammaFPS} \\
A_G &=& 1.8210(28), \; B_G = 0.435(22), 
\nonumber \\
C_G &=& -0.24(17), \; D_G = 3.2(2.9), 
\nonumber \\
E_G &=& -1.69(44), \; F_G = -69.(57.), 
\label{eq:gammaGPS} 
\end{eqnarray}
with $\chi^2/N_{df}=38.0/17$ and $\chi^2/N_{df}=19.2/17$, 
and, for $\gamma_F^*$(V) and $\gamma_G^*$(V), 
\begin{eqnarray}
A_F &=& 1.8434(48), \; B_F = 1.204(46), 
\nonumber \\
C_F &=& -0.93(29), \; D_F = -21.9(4.7), 
\nonumber \\
E_F &=& 25.94(97), \; F_F = 488.(92.), 
\label{eq:gammaFV} \\
A_G &=& 1.8311(29), \; B_G = 0.348(22), 
\nonumber \\
C_G &=& 0.14(20), \; D_G = 0.5(2.9), 
\nonumber \\
E_G &=& -1.19(47), \; F_G = -61.(58.), 
\label{eq:gammaGV} 
\end{eqnarray}
with $\chi^2/N_{df}=76.0/17$ and $\chi^2/N_{df}=14.6/17$, 
using the values of $\gamma^*_{F/G}$ and their errors listed in 
Table.~\ref{tab:calib}.
The errors for the coefficients are estimated from the $\chi^2$ error 
matrix.
These fits are shown in Figs.~\ref{fig:gamma_g} and \ref{fig:gamma_f}
by dotted lines.

In Fig.~\ref{fig:gamma_f_qmass}, we plot $\gamma_F^*$ as a function of 
the dimensionless quark mass $m_q = (1/\kappa - 1/\kappa_c)/2$ 
using $\kappa_c$ determined in Sec.~\ref{sec:physics}.
Although the range of $m_q$ is not quite close to the chiral limit, 
our values of $\gamma_F^*$ suggest a strong linear dependence in $m_q$.
This result is in clear contrast to the case of quenched QCD 
in which $\gamma_F^*$ is well fitted by a quadratic ansatz in $m_q$ 
motivated from a tree level expression of $\gamma_F^*$ 
\cite{matsufuru01}.
As mentioned in Sec.~\ref{sec:Clover}, we expect linear corrections 
from higher order quark loops even if the linear terms are 
removed at tree level \cite{aoki01}. 
Our result provides us with an example which confirms this expectation.


\begin{figure}
\resizebox{73mm}{!}{\includegraphics{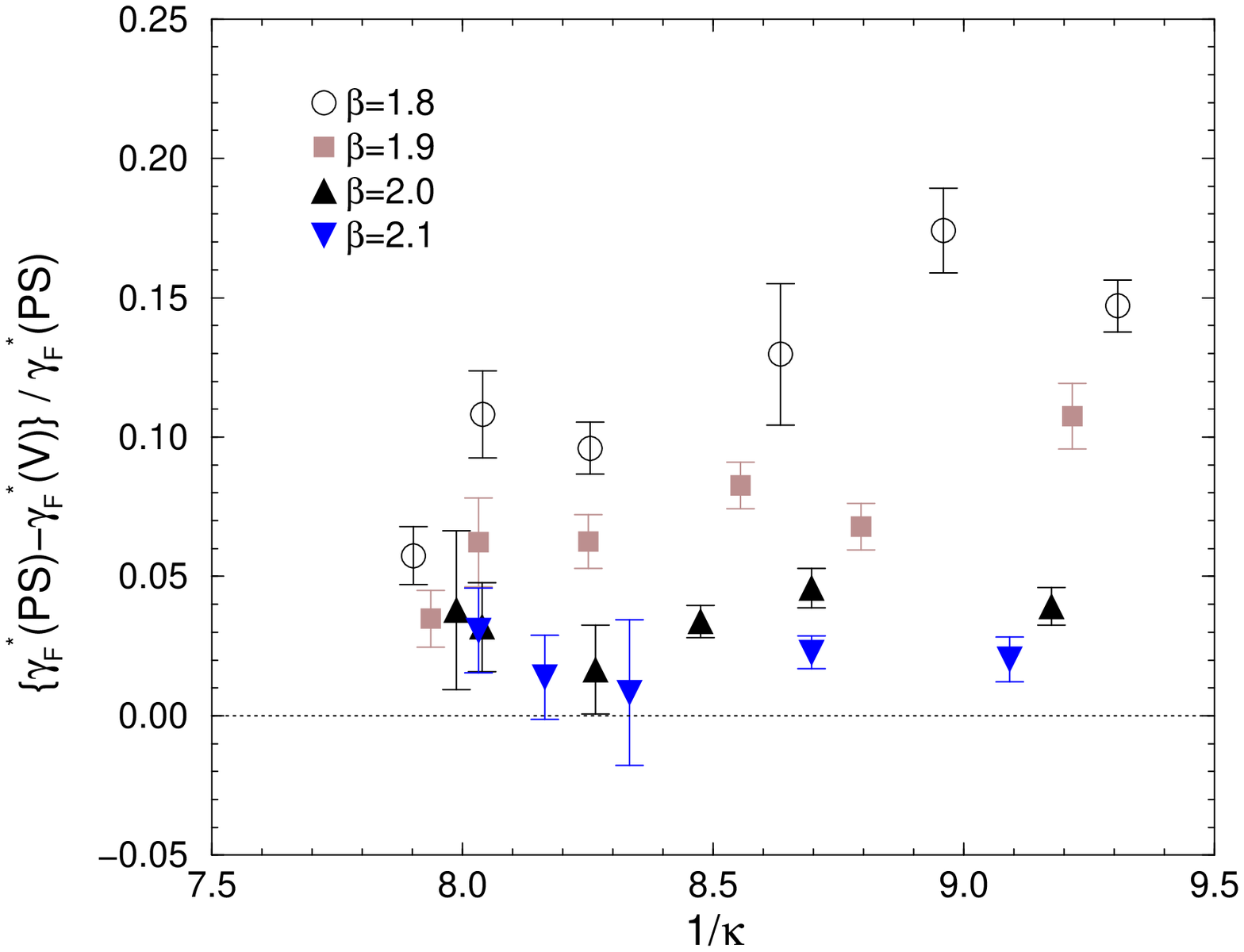}}
\resizebox{73mm}{!}{\includegraphics{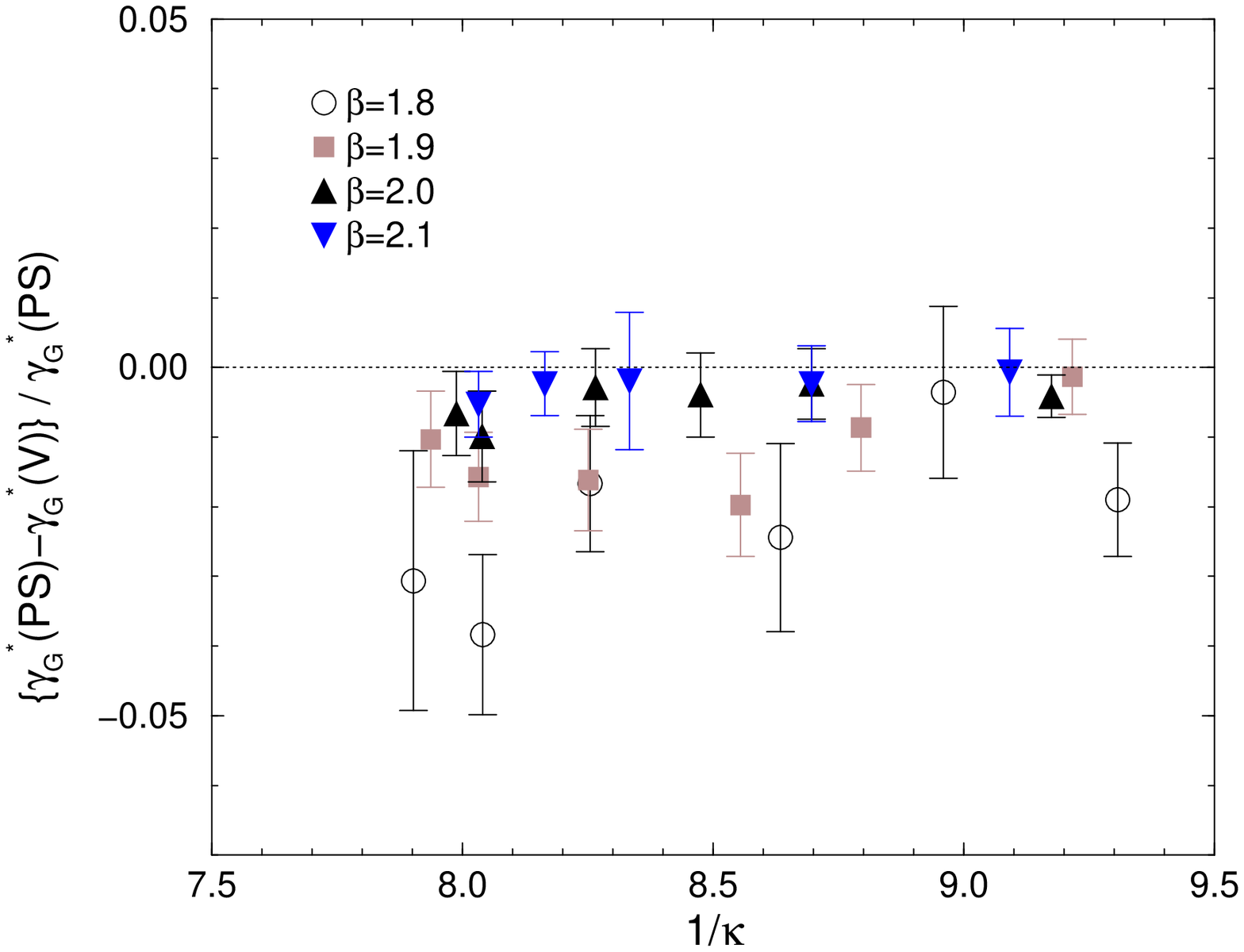}}
\caption{Relative differences 
between $\gamma_F^*({\rm PS})$ and $\gamma_F^*({\rm V})$, 
and between $\gamma_G^*({\rm PS})$ and $\gamma_G^*({\rm V})$, 
as functions of $\beta$ and $1/\kappa$.}
\label{fig:diff_psv}
\end{figure}

Finally, we study the $O(a)$ differences among the calibration results 
using PS and V mesons.
We plot relative differences
between $\gamma_F^*({\rm PS})$ and $\gamma_F^*({\rm V})$,
and between $\gamma_G^*({\rm PS})$ and $\gamma_G^*({\rm V})$, in
Fig.~\ref{fig:diff_psv} as functions of $\beta$ and $1/\kappa$.
Errors are estimated neglecting the correlation between PS and V 
determinations.
We find that the differences tend to vanish as $\beta$ is increased.
At $\beta \ge 2.0$, the differences are less than 5\% for $\gamma_F^*$
and 1\% for $\gamma_G^*$.

\section{Physical quantities at $\xi=2$}
\label{sec:physics}

\begin{table*}
\caption{Physical quantities interpolated to $\xi=2$.
Results marked by ``*'' are from the quadratic ansatz 
(\protect\ref{eq:quadansatz}), while other results are from the 
linear ansatz (\protect\ref{eq:linearansatz}).
Results of two alternatives for the $\xi=2$ point, 
$\gamma^*_{F/G}$(PS) and $\gamma^*_{F/G}$(V),
are labeled by (PS) and (V), respectively.
}
\begin{ruledtabular}
\begin{tabular}{cccccccccc}
 $\beta$ & $\kappa$ 
& $m_{PS}$(PS)     & $\chi^2/N_{df}$
& $m_V$(PS)        & $\chi^2/N_{df}$
& $m_{PS}/m_V$(PS) & $\chi^2/N_{df}$
& $r_0/a_s$(PS)    & $\chi^2/N_{df}$
\\
&
& $m_{PS}$(V)      & $\chi^2/N_{df}$
& $m_V$(V)         & $\chi^2/N_{df}$
& $m_{PS}/m_V$(V)  & $\chi^2/N_{df}$
& $r_0/a_s$(V)     & $\chi^2/N_{df}$
\\
\hline
 1.8 &  0.10745$^*$ & 
 1.4395(66) &  8.2/ 4 & 1.5760(75) &  2.6/ 4 & 0.91339(37) &  1.3/ 4 & 1.299(15)&   2.6/ 4\\
       &          & 
 1.5601(55) &  8.2/ 4 & 1.7178(64) &  2.6/ 4 & 0.90828(43) &  1.3/ 4 & 1.320(19)&   2.6/ 4\\
      &  0.11162$^*$ & 
 1.2431(62) &  1.1/ 2 & 1.3985(75) &  3.1/ 2 & 0.88900(57) &  2.8/ 2 & 1.346(18)&   1.5/ 2\\
       &          & 
 1.3009(39) &  1.1/ 2 & 1.4720(45) &  3.1/ 2 & 0.88382(65) &  2.8/ 2 & 1.355(19)&   1.5/ 2\\
      &  0.11582 & 
 1.0302(57) &  7.9/ 3 & 1.2043(78) &  7.1/ 3 & 0.8556(10) &  3.1/ 3 & 1.413(11)&   3.0/ 3\\
       &          & 
 1.079(12) &  7.9/ 3 & 1.275(17) &  7.1/ 3 & 0.8458(23) &  3.1/ 3 & 1.406(19)&   3.0/ 3\\
      &  0.12115 & 
 0.7635(15) &  2.9/ 3 & 0.9679(25) &  5.5/ 3 & 0.7888(12) &  3.0/ 3 & 1.5210(64)&   8.5/ 3\\
       &          & 
 0.7660(26) &  2.9/ 3 & 0.9906(42) &  5.5/ 3 & 0.7732(19) &  3.0/ 3 & 1.557(12)&   8.5/ 3\\
      &  0.12438$^*$ & 
 0.5880(25) &  5.7/ 4 & 0.8165(36) &  7.4/ 4 & 0.7200(22) &  6.1/ 4 & 1.605(10)&   8.1/ 4\\
       &          & 
 0.5448(73) &  5.7/ 4 & 0.7973(46) &  7.4/ 4 & 0.6834(74) &  6.1/ 4 & 1.701(13)&   8.1/ 4\\
      &  0.12655 & 
 0.4347(71) &  9.1/ 3 & 0.6841(45) &  4.3/ 3 & 0.6356(80) &  3.9/ 3 & 1.730(11)&   4.2/ 3\\
       &          & 
 0.3838(62) &  9.1/ 3 & 0.6627(46) &  4.3/ 3 & 0.5805(70) &  3.9/ 3 & 1.796(10)&   4.2/ 3\\
\hline
 1.9 &  0.10850$^*$ & 
 1.3255(56) &  4.5/ 4 & 1.4428(65) &  2.3/ 4 & 0.91851(50) &  1.5/ 4 & 1.5015(66)&  18.2/ 4\\
       &          & 
 1.4045(79) &  4.5/ 4 & 1.5330(90) &  2.3/ 4 & 0.91599(54) &  1.5/ 4 & 1.5101(73)&  18.2/ 4\\
      &  0.11370 & 
 1.0672(38) &  7.5/ 3 & 1.2002(46) &  5.7/ 3 & 0.88860(95) & 10.3/ 3 & 1.6041(80)&   2.4/ 3\\
       &          & 
 1.0977(27) &  7.5/ 3 & 1.2393(32) &  5.7/ 3 & 0.88535(59) & 10.3/ 3 & 1.6140(49)&   2.4/ 3\\
      &  0.11690 & 
 0.9085(30) &  4.9/ 5 & 1.0557(42) &  8.9/ 5 & 0.86105(98) &  9.4/ 5 & 1.6376(71)&   4.3/ 5\\
       &          & 
 0.9304(20) &  4.9/ 5 & 1.0891(27) &  8.9/ 5 & 0.85428(61) &  9.4/ 5 & 1.6742(44)&   4.3/ 5\\
      &  0.12120$^*$ & 
 0.6825(15) &  5.8/ 4 & 0.8460(27) &  5.7/ 4 & 0.8066(17) &  2.1/ 4 & 1.795(10)&   1.8/ 4\\
       &          & 
 0.6793(13) &  5.8/ 4 & 0.8516(25) &  5.7/ 4 & 0.7976(18) &  2.1/ 4 & 1.821(10)&   1.8/ 4\\
      &  0.12450$^*$ & 
 0.4859(24) &  5.9/ 3 & 0.6720(21) &  0.6/ 3 & 0.7234(28) &  2.7/ 3 & 1.9708(70)&  19.1/ 3\\
       &          & 
 0.4593(73) &  5.9/ 3 & 0.6599(40) &  0.6/ 3 & 0.6958(81) &  2.7/ 3 & 2.016(18)&  19.1/ 3\\
      &  0.12600 & 
 0.3804(61) & 33.6/ 3 & 0.5839(39) &  8.5/ 3 & 0.6517(79) & 10.7/ 3 & 2.119(14)&   1.6/ 3\\
       &          & 
 0.3559(43) & 33.6/ 3 & 0.5722(30) &  8.5/ 3 & 0.6205(56) & 10.7/ 3 & 2.165(10)&   1.6/ 3\\
\hline
 2.0 &  0.10900 & 
 1.1913(26) &  0.9/ 2 & 1.2830(30) &  1.7/ 2 & 0.92857(36) &  0.6/ 2 & 1.7792(53)&   3.9/ 2\\
       &          & 
 1.2143(35) &  0.9/ 2 & 1.3097(40) &  1.7/ 2 & 0.92725(39) &  0.6/ 2 & 1.7844(55)&   3.9/ 2\\
      &  0.11500 & 
 0.9057(18) &  2.2/ 3 & 1.0131(24) &  6.2/ 3 & 0.89420(51) &  8.3/ 3 & 1.9326(45)&   8.2/ 3\\
       &          & 
 0.9201(22) &  2.2/ 3 & 1.0323(30) &  6.2/ 3 & 0.89158(69) &  8.3/ 3 & 1.9471(66)&   8.2/ 3\\
      &  0.11800 & 
 0.75539(97) &  1.2/ 2 & 0.8696(14) &  0.3/ 2 & 0.86874(67) &  0.4/ 2 & 2.0380(48)&   5.1/ 2\\
       &          & 
 0.7604(13) &  1.2/ 2 & 0.8778(18) &  0.3/ 2 & 0.86623(92) &  0.4/ 2 & 2.0553(68)&   5.1/ 2\\
      &  0.12100 & 
 0.59609(94) & 26.3/ 3 & 0.7225(14) & 14.6/ 3 & 0.8251(15) &  0.6/ 3 & 2.229(11)&   1.5/ 3\\
       &          & 
 0.59505(95) & 26.3/ 3 & 0.7231(15) & 14.6/ 3 & 0.8229(19) &  0.6/ 3 & 2.241(11)&   1.5/ 3\\
      &  0.12440 & 
 0.3912(35) & 11.5/ 6 & 0.5397(23) & 20.9/ 6 & 0.7254(41) & 11.6/ 6 & 2.527(13)&  14.6/ 6\\
       &          & 
 0.3784(50) & 11.5/ 6 & 0.5319(31) & 20.9/ 6 & 0.7116(57) & 11.6/ 6 & 2.570(18)&  14.6/ 6\\
      &  0.12520 & 
 0.3355(37) &  8.5/ 2 & 0.4906(28) &  5.6/ 2 & 0.6843(56) &  0.1/ 2 & 2.599(16)&   0.3/ 2\\
       &          & 
 0.315(15) &  8.5/ 2 & 0.4819(67) &  5.6/ 2 & 0.655(23) &  0.1/ 2 & 2.660(47)&   0.3/ 2\\
\hline
 2.1 &  0.11000 & 
 1.0442(32) &  3.2/ 1 & 1.1150(36) &  0.2/ 1 & 0.93648(52) &  3.4/ 1 & 2.2324(72)&  13.1/ 1\\
       &          & 
 1.0545(32) &  3.2/ 1 & 1.1264(36) &  0.2/ 1 & 0.93610(47) &  3.4/ 1 & 2.2372(61)&  13.1/ 1\\
      &  0.11500 & 
 0.8108(13) &  9.0/ 2 & 0.8898(16) &  4.8/ 2 & 0.91143(49) &  0.0/ 2 & 2.4000(63)&   1.5/ 2\\
       &          & 
 0.8162(14) &  9.0/ 2 & 0.8963(17) &  4.8/ 2 & 0.91097(59) &  0.0/ 2 & 2.4134(68)&   1.5/ 2\\
      &  0.12000 & 
 0.56448(90) &  1.3/ 2 & 0.6577(14) &  1.3/ 2 & 0.8584(11) &  0.5/ 2 & 2.6992(94)&   0.3/ 2\\
       &          & 
 0.5644(14) &  1.3/ 2 & 0.6582(24) &  1.3/ 2 & 0.8577(20) &  0.5/ 2 & 2.705(18)&   0.3/ 2\\
      &  0.12250 & 
 0.4313(14) &  5.8/ 3 & 0.5339(12) &  3.6/ 3 & 0.8075(28) &  5.3/ 3 & 2.875(10)&  10.0/ 3\\
       &          & 
 0.4294(13) &  5.8/ 3 & 0.5335(13) &  3.6/ 3 & 0.8044(25) &  5.3/ 3 & 2.881(11)&  10.0/ 3\\
      &  0.12450 & 
 0.3015(23) & 10.4/ 2 & 0.4220(19) & 11.9/ 2 & 0.7134(32) &  7.9/ 2 & 3.177(15)&   1.6/ 2\\
       &          & 
 0.2906(48) & 10.4/ 2 & 0.4149(34) & 11.9/ 2 & 0.6997(63) &  7.9/ 2 & 3.229(27)&   1.6/ 2\\
\end{tabular}
\end{ruledtabular}
\label{tab:prop}
\end{table*}

In this section, we interpolate the measurement results 
to the calibration points corresponding to $\xi=2$ 
to estimate the scale and several other basic properties of our lattice. 
We also test the effects of the two calibration results 
using PS and V meson dispersion relations on the continuum extrapolation 
of physical quantities.

For the interpolations to $(\gamma_F^*,\gamma_G^*)$ 
at each $(\beta,\kappa)$, we adopt a linear ansatz
\begin{equation}
y = a + b \gamma_G + c \gamma_F
\label{eq:linearansatz}
\end{equation}
when the range of $\gamma_F$ is less than 0.3.
When $\max(\gamma_F)-\min(\gamma_F) \geq 0.3$, 
we adopt a quadratic ansatz
\begin{equation}
y = a + b \gamma_G + c \gamma_F + d \gamma_F^2,
\label{eq:quadansatz}
\end{equation}
because, in this case, the linear ansatz sometimes fails to explain 
the data ($\chi^2/N_{df} \sim O(10)$--$O(100)$). 
We find that terms quadratic in $\gamma_G$ do not improve the fits.
We confirm that this quadratic ansatz leads to a result consistent 
with the linear ansatz if $\max(\gamma_F)-\min(\gamma_F) < 0.3$.

Several physical quantities thus interpolated to $\xi=2$ are summarized 
in Table.~\ref{tab:prop},
where the results from the quadratic ansatz (\ref{eq:quadansatz}) 
are marked by ``*'' on $\kappa$.
The errors are estimated by quadratically averaging over the contributions 
from the $\chi^2$ error matrix for the fits (\ref{eq:linearansatz}) or 
(\ref{eq:quadansatz}), 
and from the errors for $\gamma_F^*$ and $\gamma_G^*$. 
Results adopting two alternative choices for the $\xi=2$ point 
--- $\gamma^*_{F/G}$(PS) from the pseudo-scalar dispersion relation 
and $\gamma^*_{F/G}$(V) from the vector dispersion relation ---
are labeled by (PS) and (V).

\subsection{Plaquette}
\label{sec:plaquette}

\begin{figure}
\resizebox{73mm}{!}{\includegraphics{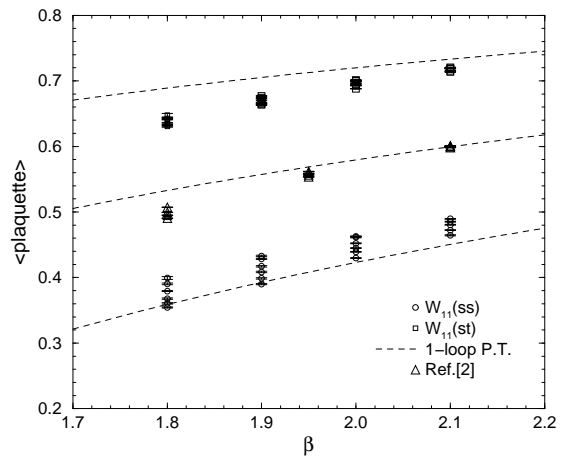}}
\caption{Plaquettes at $\xi=2$ and 1. 
Different points at the same $\beta$ with the same symbol represent 
the results obtained at different $\kappa$
(larger $W_{11}(ss)$ and $W_{11}(st)$ correspond to larger $\kappa$).
Dashed lines are the results of 1-loop perturbation theory.}
\label{fig:plaq}
\end{figure}

Figure~\ref{fig:plaq} shows the plaquette expectation values
$W_{11}(ss)$ and $W_{11}(st)$ at $\xi=2$ adopting $\gamma^*_{F/G}$(PS). 
Results adopting $\gamma^*_{F/G}$(V) are similar.
As a reference point, plaquette values on an isotropic lattice using the same 
action are also plotted \cite{cppacs02}.
Different points at the same $\beta$ with the same symbol represent 
the results obtained at different $\kappa$.

Numerical results of plaquettes are compared with their 1-loop 
values at $\xi=2$, Eqs.~(\ref{eq:wss}) and (\ref{eq:wst}), shown 
by dashed lines in the Figure. 
For $\beta$ we adopt the bare value.
We find that, as in the case of isotropic lattices, the plaquettes agree 
with the perturbative calculation within about 10\% at this value of $\beta$.
This confirms our choice of the clover coefficients $c_t$ and $c_s$ to 
this accuracy.

\subsection{Light meson spectrum and the lattice scale}
\label{sec:spectrum}

\begin{figure}
\resizebox{73mm}{!}{\includegraphics{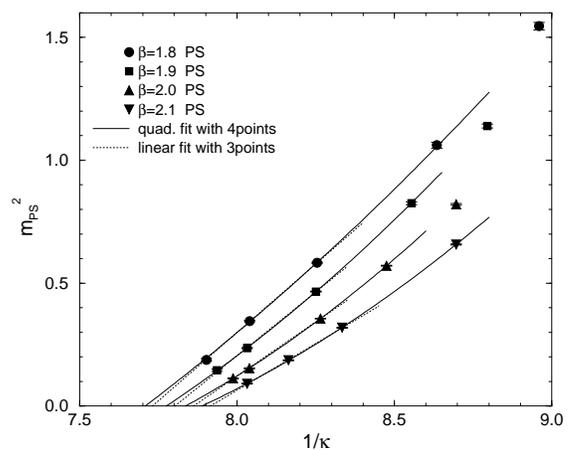}}
\caption{$m_{PS}^2$ at $\gamma^*_{F/G}$(PS) for $\xi=2$
as a function of $1/\kappa$.
Full and dotted curves show quadratic and linear fits using lightest 
four and three data points, respectively.
}
\label{fig:kappa_c}
\end{figure}

\begin{figure}
\resizebox{73mm}{!}{\includegraphics{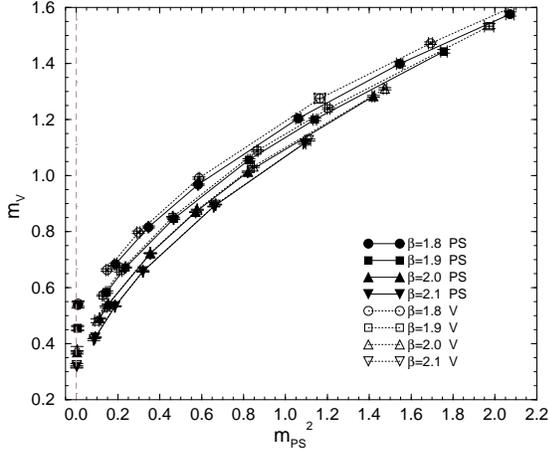}}
\caption{$m_V$ as a function of $m_{PS}^2$ for $\xi=2$
at $\gamma^*_{F/G}$(PS) (filled symbols)
and at $\gamma^*_{F/G}$(V) (open symbols).
The leftmost symbols are the results of quadratic extrapolations 
to the physical point. 
The lines are guide for the eyes.
}
\label{fig:mv}
\end{figure}

\begin{table}
\caption{Critical hopping parameter $\kappa_c$. 
The numbers in the first bracket are statistical errors, 
and those in the second bracket are systematic errors from 
the chiral extrapolation.}
\begin{ruledtabular}
\begin{tabular}{ccc}
$\beta$ & $1/\kappa_c$(PS) & $1/\kappa_c$(V) \\
\hline
1.8  & 7.708(43)($_{-0}^{+23}$)
      & 7.763(80)($_{-0}^{+25}$) \\
1.9  & 7.775(39)($_{-0}^{+27}$)
      & 7.789(46)($_{-0}^{+30}$) \\
2.0  & 7.836(26)($_{-0}^{+25}$)
      & 7.862(50)($_{-0}^{+22}$) \\
2.1  & 7.888(14)($_{-0}^{+27}$)
      & 7.896(24)($_{-0}^{+31}$) \\
\end{tabular}
\end{ruledtabular}
\label{tab:kappa_c}
\end{table}

\begin{table}
\caption{Lattice scale determined from $m_\rho$ at the physical point.}
\begin{ruledtabular}
\begin{tabular}{cccc}
$\beta$ & $a_s$(PS) [GeV$^{-1}$]  & $a_s$(V) [GeV$^{-1}$] 
& $L_s a_s$ [fm]\\
\hline
1.8  &  1.395(28)($^{+95}_{-0}$)
     &  1.408(26)($^{+67}_{-0}$)   & 2.2\\
1.9  &  1.185(21)($^{+80}_{-0}$)
     &  1.178(17)($^{+55}_{-0}$)   & 1.9\\
2.0  &  0.957(15)($^{+68}_{-0}$)
     &  0.986(26)($^{+61}_{-0}$)   & 1.9\\
2.1  &  0.824(10)($^{+62}_{-0}$)
     &  0.838(17)($^{+70}_{-0}$)   & 2.0\\
\end{tabular}
\end{ruledtabular}
\label{tab:mv}
\end{table}

In Fig.~\ref{fig:kappa_c}, we summarize the values of $m_{PS}^2$ at 
$\xi=2$ adopting $\gamma^*_{F/G}$(PS), as a function of $1/\kappa$.
Carrying out chiral extrapolations in which the 
lightest four points are fitted 
to a quadratic ansatz, we obtain the chiral point $\kappa_c$(PS) listed in 
Table~\ref{tab:kappa_c}.
The first errors are statistical, while the second ones are systematic 
errors estimated from the difference with the results of linear 
fits to the lightest three points.
The values of $\kappa_c$(V) are obtained similarly, adopting 
$\gamma^*_{F/G}$(V) as the point for $\xi=2$.
We find that $\kappa_c$(PS) and $\kappa_c$(V) are consistent with each other 
within the present statistical accuracy.

We determine the scale of our lattices from the $\rho$ meson mass 
$m_\rho=771.1$ MeV 
at the physical point $m_{PS}/m_V = m_\pi/m_\rho = 135.0/771.1$.
Figure~\ref{fig:mv} shows $m_V$ as a function of $m_{PS}^2$ at 
$\gamma^*_{F/G}$(PS) and $\gamma^*_{F/G}$(V). 
We find that the masses at $\gamma^*_{F/G}$(PS) and $\gamma^*_{F/G}$(V) 
are slightly different on coarse lattices. 
The difference rapidly decreases with increasing $\beta$.

Extrapolation to the physical point is done adopting a quadratic 
ansatz to the lightest four data points at each $\beta$.
We note that, with the present statistics, 
two results of $m_V$ at $\gamma^*_{F/G}$(PS) and $\gamma^*_{F/G}$(V), 
extrapolated to the physical point, 
are consistent with each other already on the coarsest lattice. 
High statistics simulations directly at $\gamma^*_{F/G}$(PS) and 
$\gamma^*_{F/G}$(V) may resolve the difference at the physical point. 

The resulting lattice scale is summarized in Table~\ref{tab:mv}.
Systematic errors (second errors) in the table are estimated from 
a comparison with linear fits using the lightest three points.

\subsection{Static quark potential and Sommer scale}
\label{sec:r0}

\begin{table}
\caption{Parameters for the calculation of the static quark potential.}
\begin{ruledtabular}
\begin{tabular}{ccccc}
$\beta$ & $\kappa$ & $n_{\rm smear}$ & fit-range in $t$ & fit-range in $r$ \\
\hline
1.8  & 0.10745 -- 0.11582 &  1  &  3 -- 6  &  $\sqrt{2}$ -- $2\sqrt{3}$ \\ 
     & 0.12115 -- 0.12655 &  1  &  3 -- 6  &  $\sqrt{2}$ -- $3\sqrt{2}$ \\ 
1.9  & 0.1085 -- 0.1169 &  2  &  3 -- 6  &  $\sqrt{2}$ -- $3\sqrt{2}$ \\ 
     & 0.1212 -- 0.1260 &  2  &  4 -- 7  &  $\sqrt{2}$ -- $3\sqrt{3}$ \\ 
2.0  & 0.1090 -- 0.1180 &  3  &  4 -- 7  &  $\sqrt{2}$ -- $3\sqrt{3}$ \\ 
     & 0.1210 -- 0.1252 &  3  &  5 -- 9  &  $\sqrt{3}$ -- 6.0 \\ 
2.1  & 0.1100 -- 0.1200 &  4  &  5 -- 9  &  $\sqrt{3}$ -- 6.0 \\ 
     & 0.1225 -- 0.1245 &  4  &  6 -- 11  &  2.0 -- $3\sqrt{6}$ \\ 
\end{tabular}
\end{ruledtabular}
\label{tab:potent}
\end{table}

\begin{figure}
\resizebox{73mm}{!}{\includegraphics{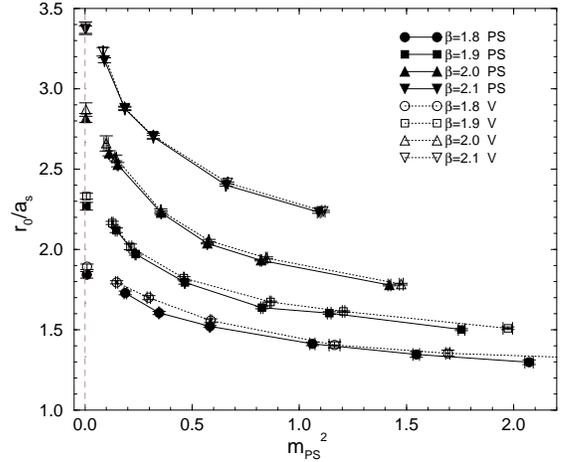}}
\caption{Sommer scale at $\xi=2$ as a function of the PS meson mass.
The leftmost symbols are the results of quadratic extrapolations 
to the physical point.
The lines are guide for the eyes.
}
\label{fig:r0}
\end{figure}

\begin{table}
\caption{Sommer scale at $\xi=2$, extrapolated to the physical point.}
\begin{ruledtabular}
\begin{tabular}{ccc}
$\beta$ & $r_0/a_s$(PS) & $r_0/a_s$(V)  \\
\hline
1.8  & 1.843(21)($^{+0}_{-47}$)
     & 1.892(17)($^{+0}_{-26}$)  \\
1.9  & 2.269(23)($^{+0}_{-77}$)
     & 2.331(18)($^{+0}_{-56}$)  \\
2.0  & 2.818(27)($^{+0}_{-62}$)
     & 2.870(41)($^{+0}_{-78}$)  \\
2.1  & 3.367(23)($^{+0}_{-91}$)
     & 3.377(39)($^{+0}_{-118}$) \\
\end{tabular}
\end{ruledtabular}
\label{tab:r0}
\end{table}

\begin{figure}
\resizebox{73mm}{!}{\includegraphics{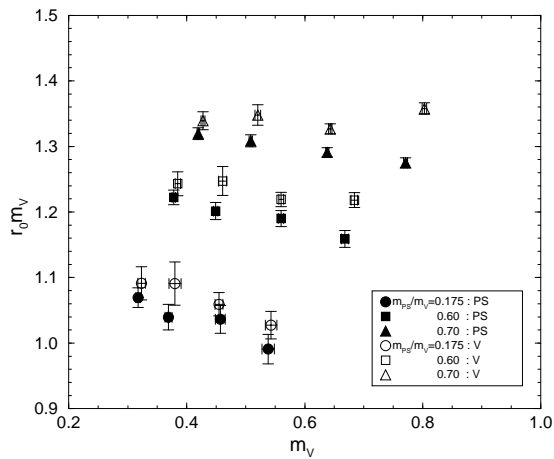}}
\caption{Sommer scale at $m_{PS}/m_V=0.7$, 0.6 and 0.175 
(the physical point) as a function of the lattice spacing. 
Errors are statistical.
}
\label{fig:cont1}
\end{figure}

\begin{figure}
\resizebox{73mm}{!}{\includegraphics{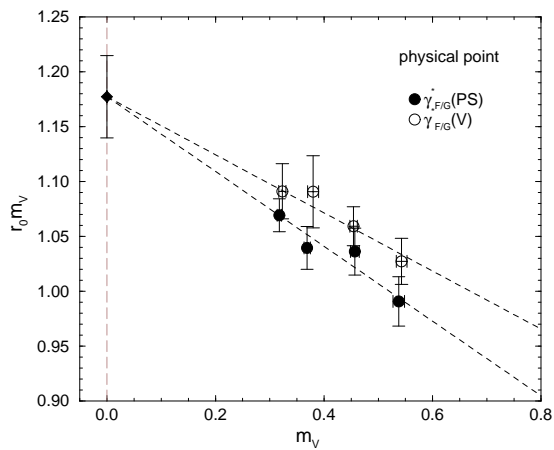}}
\caption{Sommer scale at the physical point 
as a function of the lattice spacing. 
Errors are statistical.
Lines show the continuum extrapolation with a constraint requiring the 
same continuum value for PS and V results. 
}
\label{fig:cont}
\end{figure}

We extract the static quark potential $V(r)$ from a fit
\begin{equation}
 W(r,t) = C(r) \exp{[-V(r)t]},
\end{equation}
to temporal Wilson loops. In order to enhance the overlap $C(r)$ with 
the ground state, we smear spatial links \cite{bali92}. 
We fit the data at $t\approx 0.45 - 0.90$ fm where a large overlap is 
observed.
As in previous studies on isotropic lattices, we find no apparent string 
breaking effects in $V(r)$. 
Therefore, we fit the potential with 
\begin{equation}
V(r)= A + \frac{\alpha}{r} +\sigma r. \label{eq:potent}
\label{eq:pot}
\end{equation} 
for the range $r\approx 0.35 - 1.5$ fm.
Parameters for the potential calculations are summarized in 
Table~\ref{tab:potent}.

With (\ref{eq:pot}), the Sommer scale $r_0$ \cite{sommer82} defined by 
\begin{equation}
 \left.r_0^2\frac{dV(r)}{dr}\right|_{r=r_0}=1.65
\end{equation}
is given by 
\begin{equation}
 \frac{r_0}{a_s} = \sqrt{\frac{1.65+\alpha\xi}{\sigma\xi}}.
\end{equation}
Results of $r_0/a_s$ interpolated to $\xi=2$ are listed in 
Table~\ref{tab:prop} and shown in Fig.~\ref{fig:r0}.
Extrapolating to the physical point, we obtain the values 
summarized in Table~\ref{tab:r0}, 
where the central values are from quadratic fits in $m_{PS}^2$ using 
the lightest four $\kappa$'s 
and the systematic errors are estimated from the difference with 
a linear fit using the lightest three $\kappa$'s.

The Sommer scale at $m_{PS}/m_V=0.7$, 0.6 and 0.175 
(the physical point) from a quadratic fit is plotted 
in Fig.~\ref{fig:cont1} as a function of the lattice spacing. 
We find that the difference between the calibrations 
using the PS and V meson dispersion relations 
becomes smaller toward the continuum limit.
For the Sommer scale at the physical point, 
a naive linear extrapolation to the continuum limit gives 
$r_0=0.597$(24)($^{+52}_{-14}$) and 0.612(33)($^{+79}_{-37}$) fm 
for $\gamma^*_{F/G}$ 
using the PS and V meson dispersion relations, respectively.
The systematic errors are estimated comparing the results of various 
combinations of chiral extrapolations (linear and quadratic fits for 
$m_V$ and $r_0$).
We find that the results using $\gamma^*_{F/G}$(PS) and $\gamma^*_{F/G}$(V) 
are consistent in the continuum limit within the statistical errors. 
A constrained fit requiring the same continuum value, as shown 
in Fig.~\ref{fig:cont}, 
leads to $r_0 = 0.603(19)(^{+60}_{-22})$ fm,
where the statistical error was estimated neglecting the correlation 
between the PS and V results
and the systematic error was estimated from the results of constrained 
fits using various combinations of chiral extrapolations for $m_V$ and $r_0$.

\subsection{Beta functions}
\label{sec:beta}

\begin{figure}
\resizebox{73mm}{!}{\includegraphics{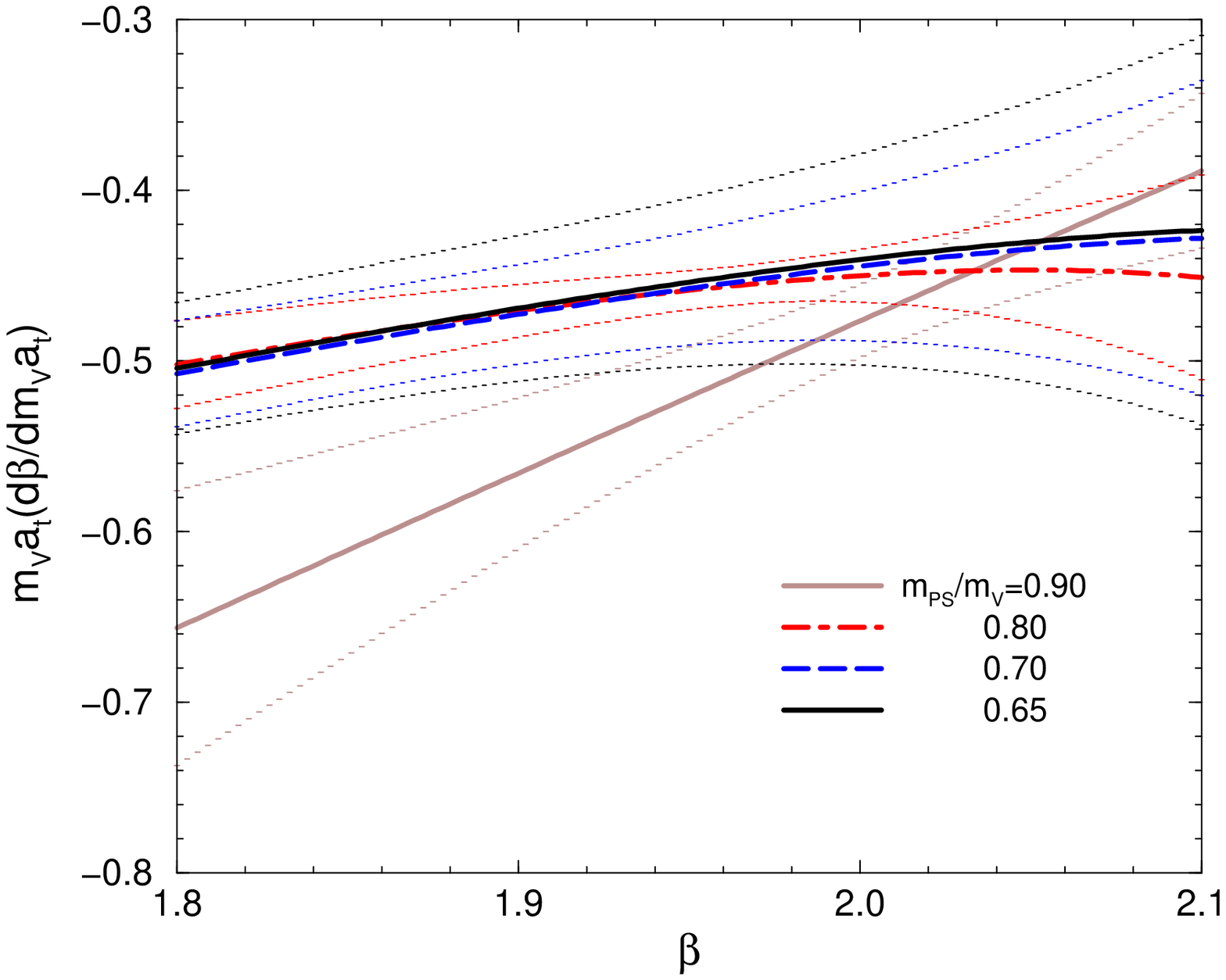}}
\resizebox{73mm}{!}{\includegraphics{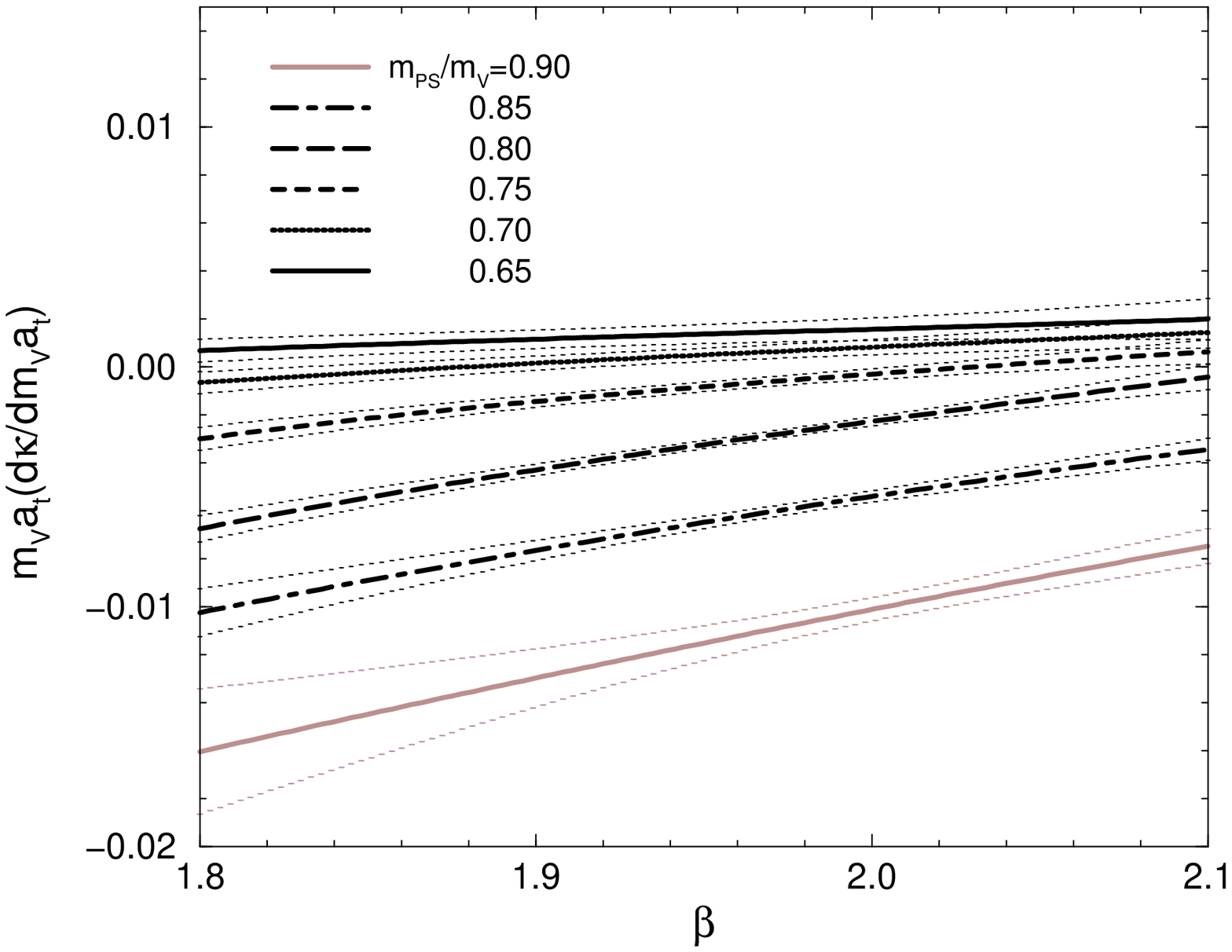}}
\caption{Beta functions (\protect\ref{eq:betafunc}) at $\gamma^*_{F/G}$(PS)
for $\xi=2$. 
Thick curves are the results for given values of $m_{PS}/m_V$, 
while thin curves represent their errors.
}
\label{fig:betaPS}
\end{figure}

\begin{figure}
\resizebox{73mm}{!}{\includegraphics{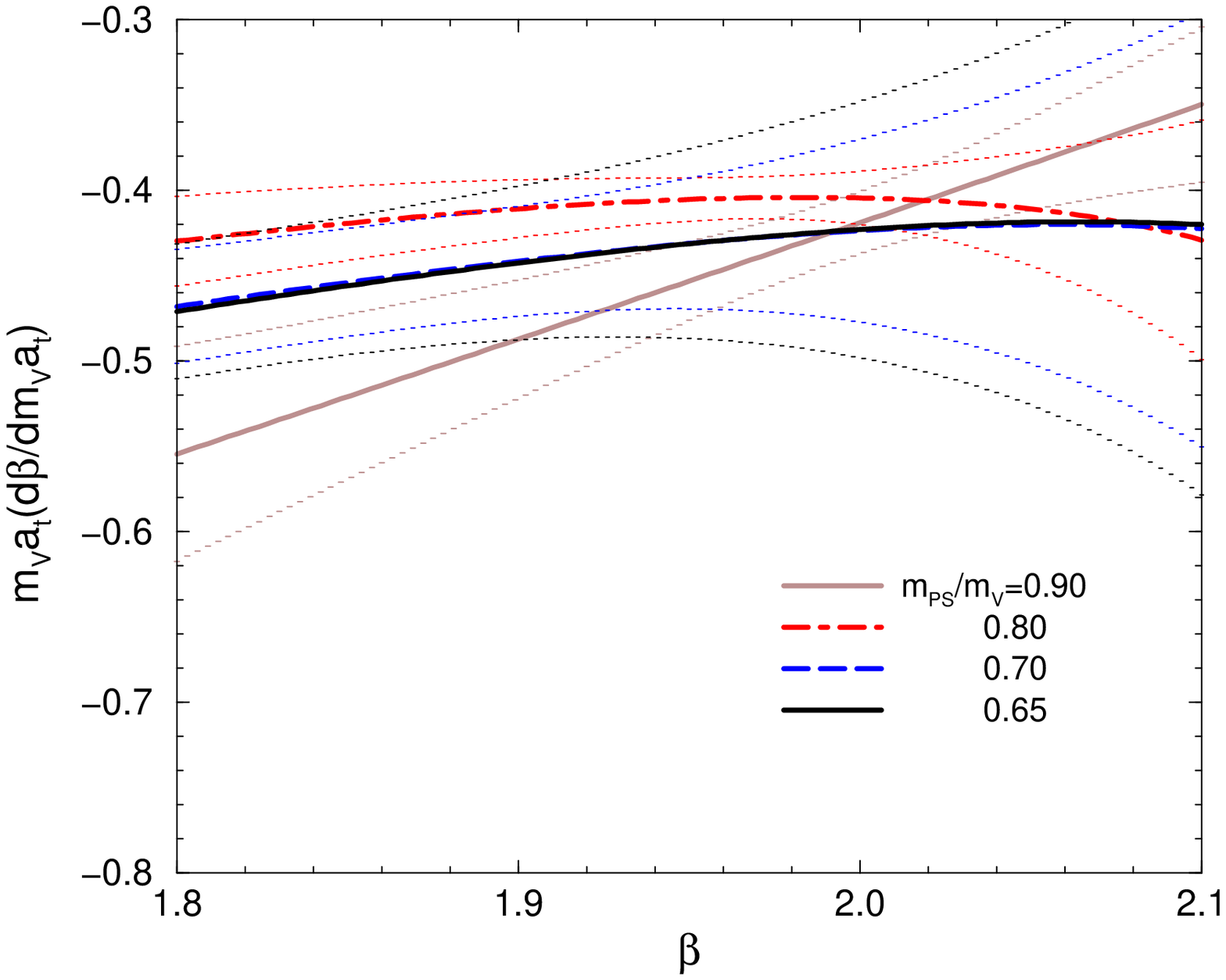}}
\resizebox{73mm}{!}{\includegraphics{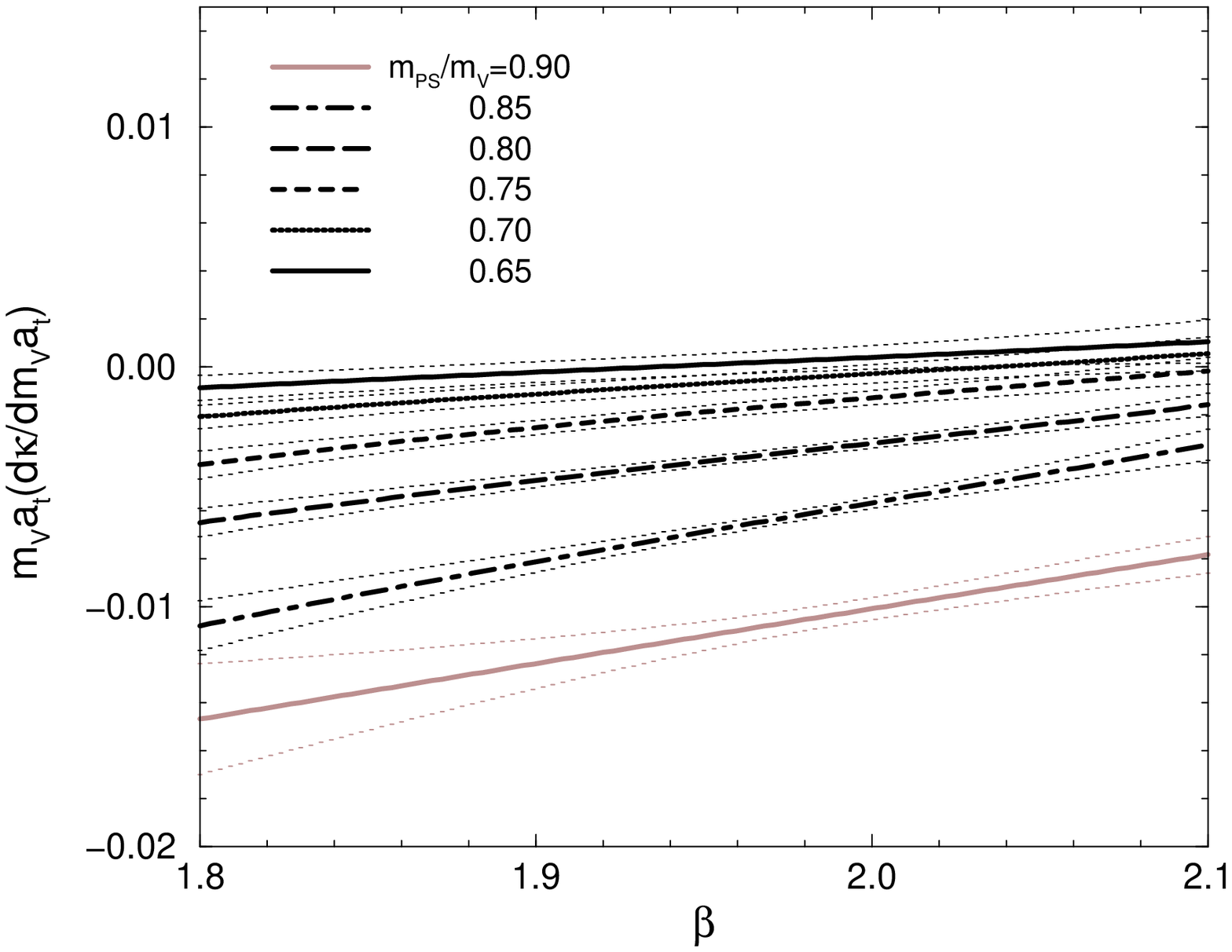}}
\caption{Same as Fig.~\protect\ref{fig:betaPS} but 
at $\gamma^*_{F/G}$(V).}
\label{fig:betaV}
\end{figure}

Finally we attempt a rough estimation of beta functions 
\begin{equation}
 a_s \left. \frac{\partial \kappa}{\partial a_s}
\right|_{\frac{m_{PS}}{m_V}}, \;\;\;
 a_s \left. \frac{\partial \beta}{\partial a_s}
\right|_{\frac{m_{PS}}{m_V}}
\label{eq:betafunc}
\end{equation}
along lines of constant physics defined by $m_{PS}/m_V =$ constant. 
These quantities are required in a calculation of the equation of state 
in thermal QCD \cite{cppacs01EOS}.
We calculate the beta functions by 
\begin{eqnarray}
&&\left( \begin{array}{cc}
\frac{\partial \beta}{\partial (m_Va_t)} &
\frac{\partial \kappa}{\partial (m_Va_t)} \\ 
\frac{\partial \beta}{\partial (m_{PS}/m_V)} &
\frac{\partial \kappa}{\partial (m_{PS}/m_V)} 
\end{array} \right)
\nonumber \\
&& =  \left( \begin{array}{cc}
 \frac{\partial (m_Va_t)}{\partial \beta} &
 \frac{\partial (m_{PS}/m_V)}{\partial \beta} \\
 \frac{\partial (m_Va_t)}{\partial \kappa} &
 \frac{\partial (m_{PS}/m_V)}{\partial \kappa} 
\end{array} \right)^{-1}
\label{eq:beta}
\end{eqnarray} 
using the data for $m_{PS}$ and $m_V$ listed in Table~\ref{tab:prop}.
In Ref.~\cite{cppacs01EOS}, a slightly different method was adopted 
because the matrix in the right hand side of (\ref{eq:beta}) 
sometimes becomes almost singular in the large quark mass region. 
Since quarks are not quite heavy in this study, we adopt the simpler 
method using (\ref{eq:beta}). 
We fit $m_{PS}$ and $m_V$ to the general quadratic ansatz in $\beta$ and 
$\kappa$.
Because the data noticeably deviates from the quadratic form when we 
include all values of $\kappa$, we restrict ourselves to three $\kappa$'s 
around the target $m_{PS}/m_V$ in the fit, while all four $\beta$'s 
are included.
Our estimates for the beta functions are summarized in 
Figs.~\ref{fig:betaPS} and \ref{fig:betaV}.

\section{Conclusions}
\label{sec:conclusion}

In this article we initiated a systematic study of two-flavor full QCD 
on anisotropic lattices. 
We have determined, for the clover-improved Wilson quarks coupled to 
an RG-improved glue, the bare anisotropy parameters 
which realize a consistent renormalized anisotropy $\xi=2$ in both 
quark and gauge sectors. In the quark sector we employed both 
pseudo scalar and vector meson channels for calibration.
The results for the bare anisotropy parameters are summarized in 
Eqs.~(\ref{eq:gammaF_form})--(\ref{eq:gammaGV})
as functions of $\beta$ and $\kappa$ for the range 
$a_s\approx 0.28$--0.16 fm and $m_{PS}/m_V \approx 0.6$--0.9.
The difference between the two calibration methods should be $O(a)$. 
We confirmed that the difference in the bare anisotropy parameters 
actually vanishes toward the continuum limit.

We have also attempted to calculate some basic quantities 
using data measured in the runs made for the calibration and interpolating 
them to the point corresponding to $\xi=2$.  
Although errors from interpolations are introduced, 
this enabled us to carry out an initial determination of 
the lattice scale and beta functions.
For the Sommer scale $r_0$, 
we found that $r_0$ from different calibration methods led to 
a consistent value in the continuum limit. 

We wish to apply our results to study of 
heavy quarks and thermal QCD, in which simulations can be directly made 
with $\xi=2$ anisotropic lattices using the parametrizations 
Eqs.~(\ref{eq:gammaF_form})--(\ref{eq:gammaGV}).   We hope to report on such 
studies in the near future.

\begin{figure}
\resizebox{73mm}{!}{\includegraphics{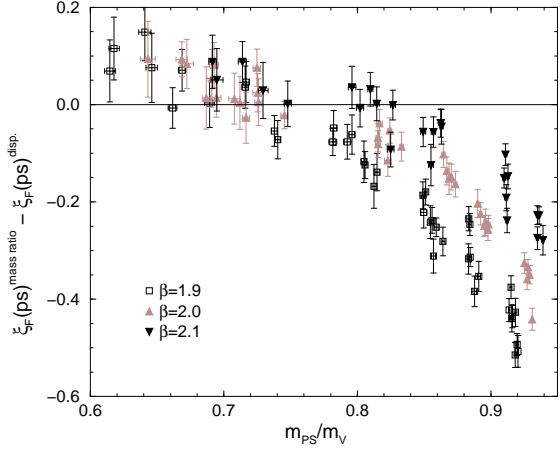}}
\caption{$\xi_F^{\rm mass \; ratio}({\rm PS})-\xi_F^{\rm disp.}({\rm PS})$
as a function of $m_{PS}/m_{V}$ at various values of $\beta$ and $\kappa$. }
\label{fig:diff_xi}
\end{figure}

\vspace*{3mm}

This work is in part supported by the Large-Scale Numerical Simulation 
Project of the Science Information Processing Center (S.I.P.C) of the
University of Tsukuba
and by the Grants-in-Aid for Scientific Research by the ministry of 
Education (Nos.~12304011, 12640253, 12740133,
13640259, 13640260, 13135204, 14046202, 11640294). 
Simulations are carried out on Fujitsu VPP5000 at S.I.P.C. 

\vspace*{5mm}

\section*{Appendix A: $\xi_F$ from the ratio of screening and temporal masses}
\label{sec:appendix}

In this paper, we adopt the dispersion relation for mesons 
for defining the fermionic anisotropy $\xi_F$. 
An alternative definition of $\xi_F$ is given by the ratio of the 
masses measures in a spatial direction (screening mass $m_s$) and 
the temporal direction (temporal mass $m_t$),
\begin{equation}
 \xi_F^{\rm mass \; ratio} = m_s/m_t,
\end{equation}
as adopted in a quenched study for PS mesons \cite{taro00}. 
For clarity, we denote $\xi_F$ defined by (\ref{eq:disp}) using 
the dispersion relation as $\xi_F^{\rm disp.}$ in this appendix.

A disadvantage of the mass ratio method is that, to obtain reliable 
values of $m_s$ and $m_t$ suppressing contamination of excited states, 
we need to prepare well-tuned smeared sources both in the spatial and 
temporal directions, and/or carry out multi-pole fits, on sufficiently 
large lattices. 
In this paper, because we do not have propagators with temporally 
smeared sources, we study spatial propagators with point-point source and sink.
We find that, when quarks are light, the effective mass of the spatial PS 
meson correlator does not show a clear plateau, and 
sometimes shows a decreasing tendency even 
at the maximum distance $x=N_s/2-1$.
This means that our spatial lattice sizes 8--12 may not be large enough to 
suppress excited states.
Unfortunately, the number of data points is also not 
sufficient to attempt a multi-pole fit. 
Therefore, in the following, we just adopt the value of effective mass 
at $x=N_s/2-1$ for $m_s$. 
Strictly speaking, this value gives an upper bound of $m_s$.
Therefore, the resulting $\xi_F^{\rm mass \; ratio}$ may be larger than 
the true value when quarks are light.

Because our temporal lattice sizes are sufficiently large, we do not 
encounter a similar problem in the calculations of $m_t$ and 
$\xi_F^{\rm disp.}$.

Figure~\ref{fig:diff_xi} shows the difference of the two fermionic 
anisotropies 
$\xi_F^{\rm mass \; ratio}({\rm PS})-\xi_F^{\rm disp.}$({\rm PS})
as a function of $m_{PS}/m_{V}$ at $\beta=1.9$, 2.0 and 2.1.
We find that the difference is consistent with zero when $m_{PS}/m_{V}$ is 
small. 
A slight overshooting at $m_{PS}/m_{V}$ \lsim 0.7 may be understood by 
the fact that our $\xi_F^{\rm mass \; ratio}$ is an upper bound for the 
true value as discussed above.
We also find that the difference at large quark masses decreases toward 
the continuum limit.
At $\beta$ \gsim 2.0 (2.1), two methods are consistent with each other at 
$m_{PS}/m_{V}$ \lsim 0.75 (0.8).

\end{document}